\documentclass[twocolumn,showpacs,preprintnumbers,amsmath,amssymb,prd,aps]{revtex4-1}
\usepackage{graphicx}
\usepackage{dcolumn}
\usepackage{bm}
\usepackage{lineno}
\usepackage{amsmath}
\usepackage{relsize}
\usepackage{color}
\usepackage{here}

\begin{document}

%\title{A generic unified model for the IceCube neutrinos and the ultrahigh-energy cosmic rays in the photomeson production scenario}

\title{Constraining photohadronic scenarios for the unified origin of \\
IceCube neutrinos and ultrahigh-energy cosmic rays}
%\title{On the unification of IceCube neutrinos and ultrahigh-energy cosmic rays\\in photohadronic scenarios}

\author{Shigeru Yoshida}
\thanks{syoshida@hepburn.s.chiba-u.ac.jp (S.~Yoshida)}
\affiliation{Department of Physics, Graduate School of Science, Chiba University, Chiba 263-8522, Japan}

\author{Kohta Murase}
\thanks{murase@psu.edu}
\affiliation{Department of Physics; Department of Astronomy \& Astrophysics; Center for Multimessenger Astrophysics, Institute for Gravitation and the Cosmos,
The Pennsylvania State University, University Park, PA 16802, USA\\
and \\
Center for Gravitational Physics, Yukawa Institute for Theoretical Physics, Kyoto University, Kyoto, Kyoto 606-8502, Japan}

\date{\today}

\begin{abstract}
  The diffuse neutrino flux measured in IceCube is comparable
  with the ultrahigh-energy cosmic ray (UHECR) flux, which has led to the concept of a unified origin of high-energy neutrino
  and UHECR backgrounds. We construct a generic unification model of sources to
  explain UHECR data at $\gtrsim 10^{19}$~eV, and high-energy neutrinos with energies that exceed $\sim100$~TeV
  in the framework of photo-meson production processes, and provide general constraints on the source properties.
  A source environment with moderately efficient in-situ production of $\gtrsim 100$~TeV neutrinos with an optical depth of
  $0.1 \lesssim \tau_{p\gamma}\lesssim 0.6$ must be realized 
  to accelerate cosmic rays to ultrahigh energies. The measured fluxes of cosmic rays and neutrinos
  set a bound on the source luminosity and its rate density. 
  Although the results are rather general and applicable to unknown source population, among the proposed source candidates,
  low-luminosity gamma-ray bursts (GRBs) and tidal disruption events (TDEs) could satisfy the requirements
  if the Lorentz bulk factor of plasma outflow and the equipartition parameters for cosmic rays and magnetic field are appropriately selected.
\end{abstract}


\maketitle

%\pagewiselinenumbers
%\linenumbers

%==================================================================================
%==================================================================================
\section{Introduction \label{sec:intro}}
%==================================================================================
%==================================================================================

The detection of cosmic neutrinos in the energy range from $\sim 10$~TeV to $\sim$~ PeV
by the IceCube Neutrino Observatory~\cite{Aartsen:2013bka,Aartsen:2013jdh,Aartsen:2014gkd,Aartsen:2016xlq}
raises interesting questions. The observed energy flux of high-energy neutrinos seems to be
comparable to that of ultrahigh-energy cosmic rays (UHECRs) at $\gtrsim 10^{19}$~eV.
Is the origin of high-energy neutrinos related to the UHECR sources?
Is the comparability of neutrino and UHECR fluxes a consequence of
yet-unknown common astrophysical phenomena? 
Several studies have been reported in the literature to probe these questions,
mainly in the framework of hadronuclear ($pp$) collisions
inside cosmic-ray reservoirs~\cite{Murase:2013rfa} -- jetted active galactic nuclei (AGN) embedded in
clusters and groups of galaxies~\cite{Fang:2017zjf}, starburst galaxies~\cite{Katz:2013ooa,Murase:2016gly},
or phenomenological setups used for UHECR observations~\cite{Kachelriess:2017tvs}. 
Remarkably, these cosmic-ray reservoir models can even explain the diffuse isotropic gamma-ray
background in the sub-TeV range, measured by the {\it Fermi} satellite~\cite{Murase:2016gly,Fang:2017zjf}. 
High-energy neutrinos can be produced, including by photohadronic interactions ($p\gamma$)
inside cosmic-ray emitters (e.g., Ref.~\cite{Winter:2013cla}). 
The photo-meson production process may occur simultaneously or in succession to the acceleration of cosmic rays. 
If the power of cosmic-ray particle emitters is sufficiently large,
it is indeed possible to emit both $\gtrsim 100$~TeV neutrinos and UHECRs.
Various astrophysical models have been investigated, which include classical high-luminosity (HL) gamma-ray bursts (GRBs)~\cite{Waxman:1997ti,Waxman:1999ai}, low-luminosity (LL) gamma-ray bursts~\cite{Murase:2006mm,Gupta:2006jm,Zhang:2018agl}, new-born magnetars~\cite{Murase:2009pg,Fang:2013vla,Fang:2018hjp}, tidal disruption events (TDEs)~\cite{Zhang:2017hom,Guepin:2017abw,Biehl:2017hnb}, and blazars~\cite{Mannheim:1995mm,Atoyan:2001ey,Essey:2009ju,Murase:2011cy,Murase:2014foa}. 

In this report, we examine a generic unification model to account for observed neutrinos
with energies greater than 100 TeV and UHECRs in the photo-meson production scheme. 
The cumulative neutrino background flux is estimated analytically using parameters to characterize sources
such as the photon luminosity and the source number density. 
The UHECR flux is also estimated semi-analytically by considering their collisions
with background photons in intergalactic space.
We also derive the source requirements for the acceleration of cosmic-ray protons
to ultrahigh energies, and transform them into the criteria of the parameters relevant
to high energy neutrino emissions such as the optical depth of $p\gamma$ interactions.
The estimated fluxes of neutrinos and UHECRs from sources that satisfy these criteria are
compared to the measured flux at 100 TeV $\lesssim E_\nu \lesssim10$~PeV and its upper limit
at $E_\nu \gtrsim100$~PeV by IceCube, as well as the measurement of UHECRs at $10^{19}$ eV.
The resultant constraints on the parameters of general source characteristics are presented. 
We finally describe a case study for specific astronomical objects such as LL GRBs.

In this work, we use $E$ for the observed energy, $\varepsilon=(1+z)E$ is the energy in the engine frame
(or the rest frame of the Hubble flow), and $\varepsilon'$ represents
the energy in the comoving frame of the plasma outflow.
The standard $\Lambda$CDM cosmology with $H_0 = 73.5$ km s$^{-1}$ Mpc$^{-1}$,
$\Omega_{\rm M} = 0.3$, and $\Omega_{\Lambda}=0.7$ is assumed throughout the report.

%==================================================================================
%==================================================================================
\section{Constraints due to Source Modeling \label{sec:model}}
%==================================================================================
%==================================================================================

%==================================================================================
%==================================================================================
\subsection{UHECR acceleration and survival \label{sec:condition}}
%==================================================================================
%==================================================================================

In the unification model of UHECRs and IceCube neutrinos, the source to emit $\gtrsim 100~{\rm TeV}$ neutrinos
must also be capable of accelerating cosmic rays to UHEs ($E_i\gtrsim 10^{20}~{\rm eV}$) by definition.
Some of the required conditions for classification as UHECR emitters are described by relatively simple formulas.

Let us consider a source with an acceleration and emission region that is given by $R$ measured in the central engine.
We also denote the bulk Lorentz factor of this source by $\Gamma$. 
For the source to account for the UHECR acceleration, the cosmic ray acceleration time scale,
$t'_{\rm acc}=\eta{\varepsilon'}_i/(Z eB^{'}c )$, must be faster than the dynamical time scale,
$t'_{\rm dyn}\approx R/(\Gamma\beta c)$. In this case, ${\varepsilon'}_i$ is
the cosmic ray ion energy in the plasma rest frame, $B'$ is the comoving magnetic field strength,
$Z$ is the atomic number of the cosmic ray ions, $\beta$ is the characteristic velocity in the source,
and $\eta^{-1}\leq1$ represents the efficiency of particle acceleration.
The condition is transformed to the well-known formula~\cite{Blandford:1999hi,Lemoine:2009pw},

%===================================
\begin{eqnarray}
L'_\gamma &\geq& \frac{1}{2}\xi_B^{-1} c\eta^2\beta^2\left(\frac{\varepsilon_i^{\rm max}}{Z e}\right)^2  \label{eq:hillas}\\
&\simeq&1.7\times 10^{45}~{\rm erg/s}~\xi_B^{-1}\eta^{2}\beta^2\left(\frac{\varepsilon_i^{\rm max}}{Z10^{11}~{\rm GeV}}\right)^2\quad\nonumber,
\end{eqnarray}
%===================================

which is equivalent to the Hillas condition in the limit of $\eta\rightarrow\beta^{-2}$.   
In this case, $\varepsilon_i^{\rm max}\approx\Gamma{\varepsilon'}_i^{\rm max}$ is the maximal energy of UHECRs accelerated at the sources. 
For a given comoving radiation luminosity $L'_\gamma$, the magnetic energy density in the plasma rest frame,
$U^{'}_{\rm B}$, is given by

%===================================
\begin{eqnarray}
U^{'}_{\rm B} &=& \xi_{\rm B}\frac{L'_\gamma}{4\pi R^2c} \nonumber\\
&=& \xi_{\rm B}\frac{L_\gamma}{4\pi \Gamma^2 R^2c}
\label{eq:magnetic_energy_density}
\end{eqnarray}
%===================================
where $\xi_{\rm B}$ is the equipartition parameter. 
For example, the modeling of GRBs and blazars typically suggests $\xi_B\sim{10}^{-4}-1$. 
As a reference value, the maximum ion energy is set to $\varepsilon_i^{\rm max}=10^{11}~{\rm GeV}$ thorough this work.
Indeed, the best fit value for the Auger data is $10^{10.9}$~GeV~\cite{Aab:2016zth},
so our choice is conservative but reasonable.
We also assume the most efficient acceleration case ($\eta=1$) and a transrelativistic or relativistic source ($\beta=1$).   

UHECRs must be accelerated before cooling via all energy loss processes
including synchrotron cooling, i.e., $t'_{\rm acc}<t'_{\rm cool}$,
where the cooling time (in the plasma rest frame) is
${t'}_{\rm cool}^{-1}={t'}_{\rm syn}^{-1}+{t'}_{p\gamma}^{-1}+{t'}_{\rm BH}^{-1}+{t'}_{\rm dyn}^{-1}$,
where $t'_{p\gamma}$ and $t'_{\rm BH}$ are the photo-meson production and Bethe-Heitler (BH) energy loss time scales, respectively. 
The last loss time scale represents adiabatic losses. For a power-law target spectrum,
the Bethe-Heitler process is important only if the spectrum is softer than
$\alpha_\gamma\sim2.2-2.3$~\cite{Murase:2018iyl}.
Therefore, we mainly consider cases wherein the BH process is subdominant.  
The synchrotron cooling time in the plasma rest frame is

%===================================
\begin{eqnarray}
{t'}_{\rm syn}^{-1} &=& \frac{4}{3}U^{'}_{\rm B}\sigma_Tc\frac{Z^4}{A^4}\frac{1}{m_pc^2}\left(\frac{\varepsilon_i}{\Gamma m_pc^2}\right)\left(\frac{m_e}{m_p}\right)^2\\
&=& \frac{4}{3}\frac{\xi_B\sigma_TL'_\gamma}{4\pi R^2}\frac{Z^4}{A^4}\frac{1}{m_pc^2}\left(\frac{\varepsilon_i}{\Gamma m_pc^2}\right)\left(\frac{m_e}{m_p}\right)^2,\nonumber
\label{eq:sync_time}
\end{eqnarray}
%===================================

where $A$ is the mass number of cosmic ray ions. 
By requiring $t'_{\rm acc}< t'_{\rm syn}$ at the maximum ion energy in the engine frame ($\varepsilon_i^{\rm max}$), we obtain

%===================================
\begin{equation}
B'<\frac{A^4 6\pi e m_p^4 c^{4}}{Z^3 \sigma_T m_e^2}\frac{\Gamma^2}{{(\varepsilon_i^{\rm max})}^2}.
\label{eq:sync_conditionpre}
\end{equation}
%===================================

In addition, we have another condition to ensure the escape of UHECRs. 
To ensure that UHECRs can leave the sources before losing their energies via synchrotron cooling,
the escape time scale $t'_{\rm esc}$ must be faster than $t'_{\rm syn}$. 
In general, the escape time is model dependent and can be long at lower energies.
For conservative estimates, we hereafter assume that the escape time scale is comparable
to the dynamical scale in the relativistic environment of the UHECR acceleration site under consideration.
This is possible if the escape boundary is comparable to the system size and
the magnetic field decays within the dynamical time (see discussion in Ref.~\cite{Zhang:2017moz}).
By regarding this ``survival'' condition as a necessary condition ($t'_{\rm dyn}<t'_{\rm syn}$), we obtain:

%===================================
\begin{equation}
B'< \frac{6\pi A^4 m_p^4 c^{9/2}}{Z^4 \sigma_T m_e^2 {(2\xi_B 
L'_\gamma)}^{1/2}}\frac{\Gamma^2}{\varepsilon_i^{\rm max}}.
\label{eq:esc_conditionpre}
\end{equation}
%===================================

We utilize Eqs.~(1), (\ref{eq:sync_conditionpre}) and (\ref{eq:esc_conditionpre}) as theoretical constraints. 
We focus on the proton case, i.e., $Z=A=1$, and discuss the cases of nuclei later.  

%==================================================================================
%==================================================================================
\subsection{Photo-meson production}
%==================================================================================
%==================================================================================

Neutrino emission by the photo-meson production process is characterized by the environment of the target photons. 
We start to build our generic framework by defining the reference energy of photons $\varepsilon_{\gamma0}$ 
in the engine frame. 
Given that a major fraction (but not necessarily all) of photo-meson production
in $p\gamma$ interactions occurs around the $\Delta$ resonance region
(including direct pion production via the $t$-channel), we introduce the reference ``resonating'' energy as

%===================================
\begin{equation}
\tilde{\varepsilon}_{p0}(s)\approx \frac{(s-m_p^2)}{4}\frac{\Gamma^2}{\varepsilon_{\gamma0}},
\label{eq:photon_energy_ref}
\end{equation}
%===================================

where $s$ is the Mandelstam variable.
In particular, we define ${\tilde{\varepsilon}}_{p0}^{\Delta}\equiv{\tilde{\varepsilon}}_{p0}(s_\Delta)$,
where $s_\Delta\approx(1.23~{\rm GeV})^2$ is the square of invariant mass of the $p\gamma$ collisions
at the $\Delta(1232)$ resonance. Primed (') characters represent
quantities measured in the rest frame of plasma with
the Lorentz bulk factor $\Gamma$. In the present model we approximate the target photon spectrum to be

%===================================
\begin{equation}
\frac{dn_\gamma}{d\varepsilon'_\gamma}= \frac{K'_\gamma}{\varepsilon'_{\gamma0}}\left(\frac{\varepsilon'_\gamma}{\varepsilon'_{\gamma0}}\right)^{-\alpha_\gamma},
 \label{eq:target_photon}
\end{equation}
%===================================

where $\alpha_\gamma$ is the photon index, where we focus on $\alpha_\gamma\geq1$. The normalization photon density $K'_\gamma$ is bolometrically connected to the source photon luminosity $L'_\gamma \approx L_\gamma/\Gamma^2$ by

%===================================
\begin{equation}
L'_{\gamma} = 4\pi R^2c\int\limits_{\varepsilon_{\gamma}^{\rm min}}^{\varepsilon_{\gamma}^{\rm max}}  \frac{dn_\gamma}{d\varepsilon'_\gamma}\varepsilon'_\gamma d\varepsilon'_\gamma.
 \label{eq:photon_energy_conversion}
\end{equation}
%===================================

We have 
%====================================
\begin{equation}
K'_\gamma = \frac{L'_{\gamma0}}{4\pi R^2 c \varepsilon'_{\gamma0}}=\left\{
\begin{array}{l}
\frac{L'_\gamma}{4\pi R^2 c \varepsilon'_{\gamma0} }\frac{\alpha_\gamma-2}{x_{\rm d}^{-\alpha_\gamma+2}-x_{\rm u}^{-\alpha_\gamma+2}} \ \ (\alpha_\gamma \neq 2)
\\
\frac{L'_\gamma}{4\pi R^2 c \varepsilon'_{\gamma0}}
\frac{1}{\ln\left(\frac{{\varepsilon'}_\gamma^{\rm max}}{{\varepsilon'}_\gamma^{\rm min}}\right)} \ \ (\alpha_\gamma =2),
\\ 
\end{array}
\right.
\label{eq:photon_luminosity}
\end{equation}
%====================================

where the two parameters
$x_{\rm d}=({\varepsilon'}_\gamma^{\rm min}/\varepsilon'_{\gamma0})$ and
$x_{\rm u}=({\varepsilon'}_\gamma^{\rm max}/{\varepsilon'}_{\gamma0})$ represents the boundary of
the main photon emission energy range that appear in the luminosity estimation, Eq.~(\ref{eq:photon_energy_conversion}).
In this case, one determines the relationship between the bolometric luminosity $L'_\gamma$ and the reference luminosity
$L'_{\gamma0}$~\footnote{This difference is known to be important for model-dependent constraints on neutrinos from GRBs}.

The optical depth to the photo-meson production is given by
(see also Eq.(6) of Ref.~\cite{Yoshida:2014uka}))

%===================================
\begin{equation}
\tau_{p\gamma}(\varepsilon'_p)=\frac{2}{1+\alpha_\gamma}\frac{L'_{\gamma0}}{4\pi R\Gamma c \varepsilon'_{\gamma0}}
\int ds \frac{\sigma_{p\gamma}(s)}{s-m_p^2}
{\left(\frac{\varepsilon'_p}{{\tilde{\varepsilon}_{p0}}^\prime(s)}\right)}^{\alpha_\gamma-1}
\label{eq:optical_depth}
\end{equation}
%===================================

where $\varepsilon'_p\approx\varepsilon_p/\Gamma$ and $\sigma_{p\gamma}$ is the photo-meson production cross-section. 
Using the approximation $\sigma_{p\gamma}\approx(s_\Delta-m_p^2)\bar{\sigma}_{\Delta}\delta(s-s_{\Delta})$
(where $\bar{\sigma}_{\Delta}\sim3\times{10}^{-28}~{\rm cm}^2$ is the cross-section averaged over the resonance range),
we reproduce the known results (e.g., Refs.~\cite{Waxman:1997ti,Dermer:2012rg,Murase:2015xka}
with inelasiticity taken into account). 
Using this resonance approximation, the preceding equation is rewritten as

%===================================
\begin{equation}
\tau_{p\gamma}(\varepsilon_p)\approx\frac{2}{1+\alpha_\gamma}\frac{L'_{\gamma0}}{4\pi R\Gamma^2 c (\varepsilon'_{\gamma0}/\Gamma)} {\left(\frac{\varepsilon_p}{{\tilde{\varepsilon}_{p0}^{\Delta}}}\right)}^{\alpha_\gamma-1}
\int ds \frac{\sigma_{p\gamma}(s)}{s-m_p^2}
\label{eq:optical_depth2}
\end{equation}
%===================================

This approximation is valid for $\alpha_\gamma\gtrsim1$, and for $\alpha_\gamma\sim1$ there
is an enhancement by a factor of $2-3$ due to multipion production~\cite{Murase:2005hy}. 
Since $E_\nu\sim1$~PeV neutrinos originate from $\varepsilon_p\approx(1+z)20$~PeV protons
(where $\bar{z}$ is the typical source redshift), we use $\tilde{\varepsilon}_{p0}^{\Delta}$
as the reference proton energy (in the engine frame), which is fixed to $\tilde{\varepsilon}_{p0}^{\Delta}=10$~PeV.
In this case, one should consider that this implicitly requires target photons that can resonantly interact
with protons with an energy of 10~PeV. As such, $\varepsilon'_{\gamma0}$ has an implicit $\Gamma$ dependence
via Eq.~(\ref{eq:photon_energy_ref}). We have

%===========================================================
\begin{equation}
\varepsilon_{\gamma0}\approx16~\Gamma^2{(\tilde{\varepsilon}_{p0}^{\Delta}/10~{\rm PeV})}^{-1}~{\rm eV}.
\label{eq:resnuenergy}
\end{equation}
%===========================================================

As a result, one can see from Eq.~\ref{eq:optical_depth2} that
$\tau_{p\gamma}(\tilde{\varepsilon}_{p0}^{\Delta})\equiv \tau_{p\gamma0}\propto L_\gamma \Gamma^{-2}{(\varepsilon_{\gamma0})}^{-1}R^{-1}\propto L'_\gamma \Gamma^{-1}{(\varepsilon'_{\gamma0})}^{-1}R^{-1}\propto L'_\gamma \Gamma^{-2}R^{-1}\tilde{\varepsilon}_{p0}^{\Delta}$.

%============================x==============
\begin{figure}
  \begin{center}
  \includegraphics[width=0.45\textwidth]{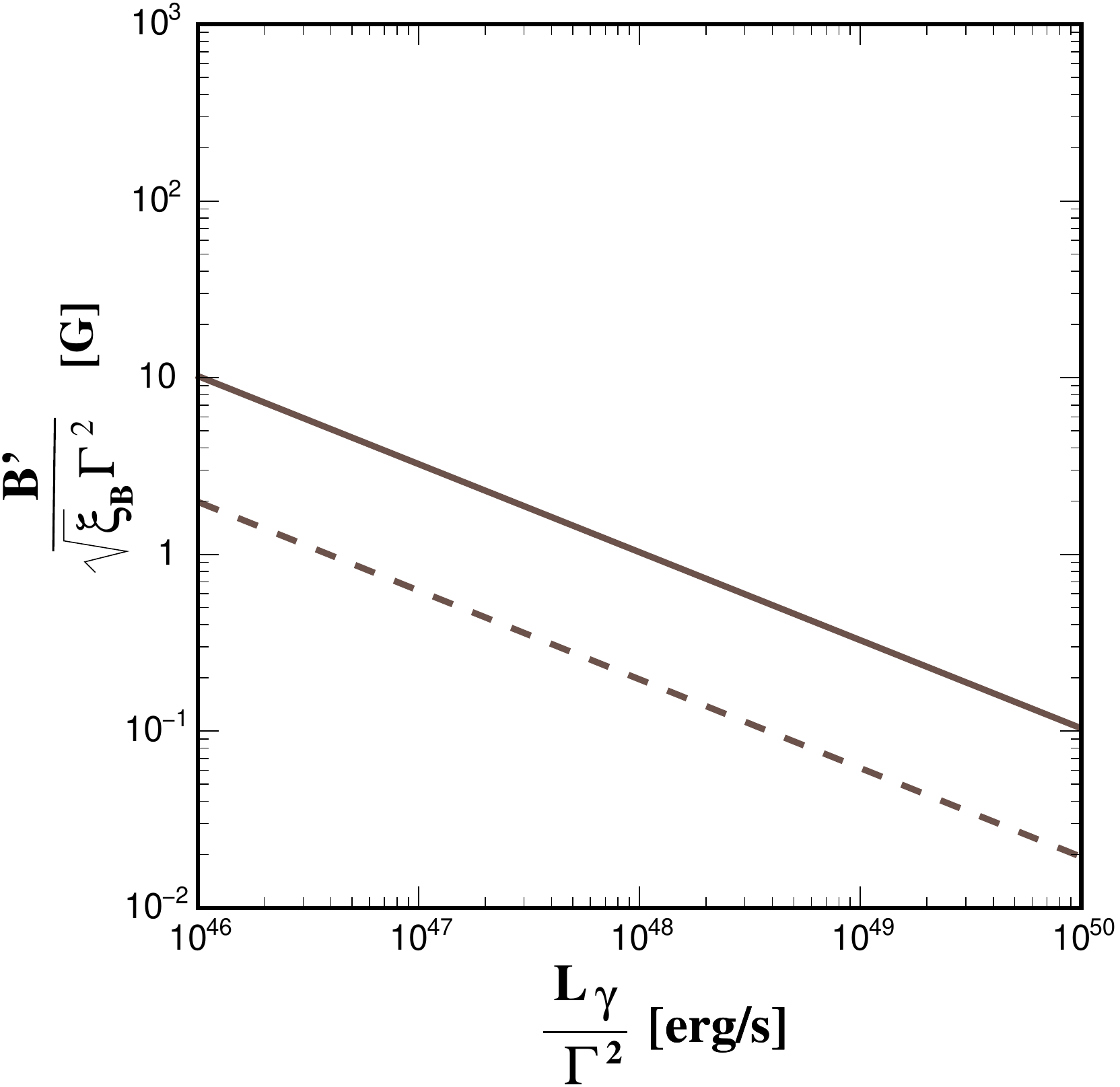}
  \caption{The relationship between the comoving magnetic field strength ${\rm B'}$ and the comoving photon luminosity
    $L'_\gamma\approx L_\gamma/\Gamma^2$. The solid line displays the case
    when $1-\exp{(-\tau_{p\gamma0})}=0.4$ (corresponding to $\tau_{p\gamma0}\sim1$),
    which is the most optically thick case allowed by the UHECR escape condition.
    The dashed line shows the $B'-L'_\gamma$ relationship when $1-\exp{(-\tau_{p\gamma0}})=0.1$
    (corresponding to $\tau_{p\gamma0}\sim0.1$).}
  \label{fig:optical_depth}
  \end{center}
\end{figure}
%==========================================

The emission radius $R$ appears in Eq.~(\ref{eq:optical_depth2}), but it can be eliminated via Eq.~(2).
For a given value of $\tilde{\varepsilon}_{p0}^{\Delta}$,
as $R\propto L'_\gamma/\tau_{p\gamma0}/\Gamma^2$ and $U^{'}_{\rm B}\propto \xi_B L'_\gamma/R^2$,
the magnetic field strength must satisfy

%====================================
\begin{equation}
\frac{B'/\Gamma^2}{\tau_{p\gamma 0}\sqrt{\xi_B/ L'_\gamma}}={C(\alpha_\gamma,\tilde{\varepsilon}_{p0}^{\Delta})}^{-1},
\label{eq:optical_depth3}
\end{equation}
%====================================

where $C(\alpha_\gamma,\tilde{\varepsilon}_{p0}^{\Delta})$ is a constant that depends
on the photon index. For $\bar{\sigma}_{\Delta}\sim3\times{10}^{-28}~{\rm cm}^2$, we have

%====================================
\begin{eqnarray}
C(\alpha_\gamma,\tilde{\varepsilon}_{p0}^{\Delta})&\sim&2.4\times{10}^{-24}~{\rm erg}^{-1}~{\rm cm}^{3/2}~{\rm s}^{1/2}\nonumber\\
&\times&\left(\frac{2}{1+\alpha_\gamma}\right)\left(\frac{\tilde{\varepsilon}_{p0}^{\Delta}}{10~{\rm PeV}}\right)\left(\frac{5L'_{\gamma0}}{L'_\gamma}\right).
\end{eqnarray}
%====================================

The source model has been constructed such that for a given $L'_\gamma$ and $\Gamma$,
the $p\gamma$ interaction site radius $R$ can arbitrarily vary to realize various values of
$\tau_{p\gamma0}$ and $B'$ (assuming a value of the equipartition parameter $\xi_B$)
via Eqs.~(\ref{eq:optical_depth2}) and (\ref{eq:magnetic_energy_density}).
This enables us to eliminate the model dependence on $R$ that is often very uncertain
(see Refs.~\cite{Murase:2005hy,Murase:2008mr} for GRBs).
Eq.~(\ref{eq:optical_depth3}) can further be combined with the conditions for UHECR acceleration and survival. 
The explicit independent parameters for this construction are then
$L'_\gamma$, $\Gamma$, and  $\tau_{p\gamma0}$, as well as the subparameters $\xi_B$ and $\alpha_\gamma$.

With Eqs.~(\ref{eq:sync_conditionpre}) and (\ref{eq:optical_depth3}),
one of the UHECR acceleration conditions gives the following upper limit
on the $p\gamma$ optical depth:

%====================================
\begin{equation}
%\begin{array}{l}
\tau_{p\gamma0}<\frac{C(\alpha_\gamma,\tilde{\varepsilon}_{p0}^{\Delta})6\pi e m_p^4 c^4}{\sigma_T m_e^2}\frac{A^4}{Z^3}\left(\frac{{L'}_\gamma^{1/2}}{\xi_B^{1/2}{(\varepsilon_i^{\rm max})}^2}\right).
\\
%\end{array}
\label{eq:sync_condition}
\end{equation}
%====================================

%%===================================
%%===================================
%\begin{equation}
%\begin{array}{l}
%\frac{6e}{\sigma_T\tau_{p\gamma0}}\frac{\alpha_\gamma-2}{\alpha_\gamma+1}\sqrt{2\frac{L'_\gamma}{\xi_{\rm B}c}} \left(\frac{m_p}{m_e}\right)^2(s_R-m_p^2)^{\alpha_\gamma-2}\frac{E_o^s}{x_{\rm d}^{-\alpha_\gamma+2}-x_{\rm u}^{-\alpha_\gamma+2}}\\
%\int ds \sigma_{\gamma p}(s)(s-m_p^2)^{-\alpha_\gamma}\geq \left(\frac{E_i^{\rm max}}{m_p %c^2}\right)^2. \\
%\end{array}
%\label{eq:sync_condition}
%\end{equation}
%%===================================

By applying the UHECR escape condition (\ref{eq:esc_conditionpre}) to  the optical depth formula,
Eq.~(\ref{eq:optical_depth2}), we can obtain the condition for $\tau_{p\gamma0}$
without explicitly depending on $L'_{\gamma}$ and $\Gamma$ as:

%===================================
\begin{eqnarray}
\tau_{p\gamma0}&<& \frac{2}{1+\alpha_\gamma}\left(\int ds \frac{\sigma_{p\gamma}}{s-m_p^2}\right)\frac{3A^4m_p^4c^4(L'_{\gamma0}/L'_{\gamma})}{4Z^4\sigma_Tm_e^2(\varepsilon'_{\gamma0}/\Gamma)}\frac{1}{\xi_B\varepsilon_i^{\rm max}} \nonumber \\
&\lesssim& 6\times 10^{-2} \frac{2}{1+\alpha_\gamma}\xi_B^{-1}{\left(\frac{A}{Z}\right)}^4{\left(\frac{\varepsilon_i^{\rm max}}{10^{11}\ {\rm GeV}}\right)}^{-1}
%&\lesssim& 4.1\times 10^{-2}\left(\frac{E_i^{\rm max}}{2\times 10^{11}\ {\rm GeV}}\right)^{-1}\xi_{\rm B}^{-1}
\label{eq:esc_condition}
\end{eqnarray}
%===================================
%%===================================
%\begin{widetext}
%\begin{eqnarray}
%  \tau_{p\gamma0} &\leq & \frac{6}{\xi_{\rm %B}\sigma_T}\frac{\alpha_\gamma-2}{\alpha_\gamma+1}\left(\frac{m_p}{m_e}\right)^2\left(\frac{%E_i^{\rm max}}{m_p c^2}\right)^{-1} \
%  \frac{E_o^s m_p c^2}{x_{\rm d}^{-\alpha_\gamma+2}-x_{\rm %u}^{-\alpha_\gamma+2}}(s_R-m_p^2)^{\alpha_\gamma-2} \
%  \int ds \sigma_{\gamma p}(s)(s-m_p^2)^{-\alpha_\gamma} \nonumber\\
%  &\lesssim& 4.1\times 10^{-2}\left(\frac{E_i^{\rm max}}{2\times 10^{11}\ {\rm %GeV}}\right)^{-1}\xi_{\rm B}^{-1}
% \label{eq:esc_condition}
%\end{eqnarray}
%\end{widetext}
%%===================================
It should be noted that $\varepsilon'_{\gamma0}/\Gamma = (s_\Delta-m_p^2)/4\tilde{\varepsilon}_{p0}^\Delta$ and
thus, this bound is $\Gamma$ independent.

Fig.~\ref{fig:optical_depth} displays this dependence for two representative cases
for $\tau_{p\gamma0}$ -- the optical depth of protons with energy $\tilde{\varepsilon}_{p0}^{\Delta}=10$~PeV.

It should be noted that the UHECR acceleration and survival conditions require that
$t'_{\rm acc}<t'_{p\gamma}$ and $t'_{\rm dyn}<t'_{p\gamma}$ should also be satisfied.
The latter condition means that the system should not be calorimetric to the sources
to simultaneously account for the IceCube neutrino and UHECR fluxes. Therefore, we have

%===================================
\begin{equation}
\tau_{p\gamma0}\leq\tau_{p\gamma}(\varepsilon_p^{\rm max})\lesssim 1/\kappa_{p\gamma}\sim5, 
\label{eq:calorimetric}
\end{equation}
%===================================

where $\kappa_{p\gamma}\sim0.2$ is the proton inelasticity.
In the cases whereby $\alpha_{\rm cr}\geq2$, this condition is satisfied
by the diffuse flux measurements (see below), so that $t_{\rm acc}<t_{\rm dyn}<t_{p\gamma}$ is automatically fulfilled.

%==================================================================================
%==================================================================================
%==================================================================================
\section{Constraints due to Diffuse UHECR and Neutrino Fluxes \label{sec:diffuse}}
%==================================================================================
%==================================================================================
%==================================================================================

An important observation is that the energy generation rate densities of UHECRs and neutrinos
are comparable~\cite{Katz:2013ooa,Murase:2018utn}. The detailed comparison of these fluxes
constrains the parameter space of the unification model. 

%==================================================================================
%==================================================================================
\subsection{\label{sec:neutrinoFlux} Neutrino spectra with radiative cooling of mesons and muons}
%==================================================================================
%==================================================================================

The flux of high-energy neutrinos for a given optical depth $\tau_{p\gamma0}$ has
been calculated using various analytical and numerical methods.
In this work, based on Ref.~\cite{Yoshida:2014uka}, we outline the analytical formulation
and its minor modifications to account for the synchrotron cooling of mesons and muons.

The spectrum of UHECRs injected from the UHECR sources is assumed to follow a power-law form, which is
%==================================================================================
%\begin{equation}
%  \frac{d\dot{N}_{\rm CR}}{d\varepsilon_i}=\frac{K_{\rm CR}}{\varepsilon_{i0}}\left(\frac{\varepsilon_i}{\varepsilon_{i0}}\right)^{-\alpha_{\rm CR}}.
%  \label{eq:UHECRspec}
%\end{equation}
%==================================================================================
%===================================
\begin{equation}
\frac{d\dot{N}_{\rm CR}}{d\varepsilon_i}=\frac{K_{\rm CR}}{\varepsilon_{i0}}\left(\frac{\varepsilon_i}{\varepsilon_{i0}}\right)^{-\alpha_{\rm CR}}e^{-\varepsilon_i/\varepsilon_i^{\rm max}},
\label{eq:UHECRspec}
\end{equation}
%===================================

where $\varepsilon_{i0}$ is the reference energy that can be set to $\tilde{\varepsilon}^\Delta_{p0}$ for protons.
The normalization factor, $K_{\rm CR}$, of the UHECR yield (with a dimension of [s]$^{-1}$)
is linked quasi-bolometrically to the photon luminosity $L_\gamma$ with the CR loading factor $\xi_{\rm CR}$:

%===================================
\begin{equation}
K_{\rm CR}\approx\left\{
\begin{array}{l}
\frac{(\alpha_{\rm CR}-2)\xi_{\rm CR}L_{\gamma}/\varepsilon_{i0}}{(\frac{\varepsilon_i^{\rm min}}{\varepsilon_{i0}})^{-\alpha_{\rm CR}+2}-(\frac{\varepsilon_i^{\rm max}}{\varepsilon_{i0}})^{-\alpha_{\rm CR}+2}} \quad  (\alpha_{\rm CR} \neq 2) \\
\\
\frac{\xi_{\rm CR}L_\gamma/\varepsilon_{i0}}{\ln\left(\frac{\varepsilon_i^{\rm max}}{\varepsilon_i^{\rm min}}\right)} \quad (\alpha_{\rm CR} = 2). 
\end{array}
\right.
\label{eq:uhecr_luminosity}
\end{equation}
%===================================

Assuming that UHECRs are protons, we set $\varepsilon_i^{\rm min}=\tilde{\varepsilon}^\Delta_{p0}=10$~PeV hereafter.

In general, if pions and muons decay into gamma rays and leptons without energy loss,
the differential neutrino luminosity from a single source, $d\dot{N}_{\nu}/d\varepsilon_{\nu}$ is
formally given by~\cite{Murase:2007yt,Murase:2014foa}

%===================================
\begin{equation}
\frac{d\dot{N}_{\nu}}{d\varepsilon_{\nu}}
\approx \int d\varepsilon_{i} \frac{d\dot{N}_{\rm CR}}{d\varepsilon_i} \int d\varepsilon'_\gamma \frac{dn_\gamma}{d\varepsilon'_\gamma} 
\left\langle \frac{d \sigma_{p\gamma\rightarrow\nu}}{d\varepsilon_\nu}(\varepsilon_i,\varepsilon'_\gamma) \right\rangle c t'_{\rm cool},
\label{eq:general_yield}
\end{equation}
%===================================

where $d\sigma_{p\gamma\rightarrow\nu}/d\varepsilon_\nu$ is the inclusive differential cross-section
with the multiplicity of neutrinos taken into account.
Given that we focus on $t'_{\rm cool}\approx t'_{\rm dyn}$, with the energy dependent optical depth
$\tau_{p\gamma}\approx \int d\varepsilon'_\gamma (dn_\gamma/d\varepsilon'_\gamma)\langle \sigma_{p\gamma} \rangle ct'_{\rm dyn}$,
the preceding equation is approximated to be~\cite{Yoshida:2014uka}

%===================================
\begin{equation}
\frac{d\dot{N}_{\nu}}{d\varepsilon_{\nu}}\approx  
\int d\varepsilon_i \frac{K_{\rm CR}}{\varepsilon_{i0}} \left( \frac{\varepsilon_i}{\varepsilon_{i0}} \right)^{-\alpha_{\rm CR}}
Y(\varepsilon_{\nu};\varepsilon_i) \tau_{p\gamma}(\varepsilon_i).
\label{eq:general_yield_approx}
\end{equation}
%===================================

In this case, $Y(\varepsilon_{\nu};\varepsilon_i)$ denotes the energy distribution of the neutrinos produced
by an interaction of a cosmic ray proton. The details of the expression for $Y$ are given in Appendix A
(see also Refs.~\cite{He:2012tq,Kimura:2017kan} for another analytical approximation).
It should be noted that the neutrino spectrum cannot be harder than $\propto \varepsilon_\nu^0$~\cite{Gaisser:1990vg,Murase:2015xka}.

The radiative cooling of pions and muons is important when the cooling time
becomes ``shorter'' than the decay time~\cite{Waxman:1997ti,Razzaque:2004yv}, provided that
their escape time from the turbulent magnetic field region is much longer.
In general, various processes such as inverse Compton and adiabatic losses can be relevant.
We consider the case of synchrotron dominance.
The ratio of the synchrotron cooling time to the decay time can be written as

%===================================
\begin{equation}
\frac{t'_{\pi/\mu,{\rm syn}}}{t'_{\pi/\mu,\rm dec}}=\left(\frac{\varepsilon_{\nu,\pi/\mu}^{\rm syn}}{\varepsilon_\nu}\right)^2,
\label{eq:synchrotron_factor}
\end{equation}
%===================================

where $t'_{\pi/\mu,{\rm syn}}$ is the synchrotron time scale of pions (muons),
$t'_{\pi/\mu,{\rm dec}}=[\varepsilon'_{\pi/\mu}/(m_{\pi/\mu}c^2)]\tau_{\pi/\mu}$
is the lifetime of pions and muons, and $\tau_{\pi/\mu}$ is their proper lifetime.
In addition, $\varepsilon_{\nu,\pi/\mu}^{\rm syn}$ is the critical neutrino energy of a pion (or muon),
above which the suppression due to synchrotron cooling is relevant.
The critical energy is given by~\cite{Waxman:1997ti,Murase:2011cx}

%==================================================================================
\begin{equation}
\varepsilon_{\nu,\pi/\mu}^{\rm syn} \approx \Gamma\kappa_{\pi,\mu} \sqrt{\frac{6\pi}{\tau_{\pi,\mu}\sigma_TcB'^2}\frac{(m_{\pi/\mu}c^2)^5}{(m_ec^2)^2}},
\label{eq:critical_synchrotron_energy}
\end{equation}
%==================================================================================

where $\kappa_{\pi,\mu}$ is the inelasticity from pion (muon) to a neutrino in the decay process.
In this work, $\kappa_{\pi}$ is approximated by $\sim 1-r_{\pi}$~\cite{Gaisser:1990vg},
where $r_{\pi} = m_{\mu}^2 / m_{\pi}^2 \simeq 0.57$ is the muon-to-pion mass-squared ratio.
The other fraction goes to a muon and $\kappa_\mu$ is approximated as $\sim 0.3$.

In the synchrotron cooling energy regime, {\it i.e.,},
$\varepsilon_\nu\geq\varepsilon_{\nu}^{\rm syn}$,
the neutrino yield is suppressed by $t'_{\pi/\mu,\rm syn}/t'_{\pi/\mu,\rm dec}$. 
Introducing $f_{\rm sup}(\varepsilon_\nu)=1-\exp(-t'_{\pi/\mu,{\rm syn}}/t'_{\pi/\mu,{\rm dec}})$~\cite{He:2012tq,Kimura:2017kan},
the neutrino yield, Eq.~(\ref{eq:general_yield}), is modified as

%===================================
\begin{eqnarray}
\frac{d\dot{N}_{\nu}}{d\varepsilon_{\nu}}\approx  
\int d\varepsilon_i \frac{K_{\rm CR}}{\varepsilon_{i0}} \left( \frac{\varepsilon_i}{\varepsilon_{i0}} \right)^{-\alpha_{\rm CR}}
\tau_{p\gamma}(\varepsilon_i)Y(\varepsilon_{\nu};\varepsilon_i)f_{\rm sup}(\varepsilon_\nu).\,\,\,\,\,\,\,\,\,
\label{eq:general_yield_wz_sync}
\end{eqnarray}
%===================================

The break energy of the neutrino flux due to synchrotron cooling is 
given by Eq.(\ref{eq:critical_synchrotron_energy}) and scales as $\varepsilon_{\nu}^{\rm syn}\propto \Gamma R/\sqrt{L'_\gamma}$.
Since the optical depth $\tau_{p\gamma0}$ scales as $\propto L'_\gamma/(R\Gamma^2)$,
we get $\varepsilon_{\nu}^{\rm syn}\propto \sqrt{L'_\gamma}/(\Gamma\tau_{p\gamma0})$.
Thus, the upper limit of the neutrino flux in the energy region beyond 10 PeV for IceCube~\cite{Aartsen:2018vtx}
can constrain $L'_\gamma$, $\Gamma$ and $\tau_{p\gamma0}$.

%
%==================================================================================
%==================================================================================
\subsection{Calculations of diffuse intensities}
%==================================================================================
%==================================================================================

Assuming emission from standard candles (i.e., identical sources over redshifts),
the energy flux of diffuse neutrinos from UHECR sources across the universe,
$\Phi_\nu\equiv dJ_{\nu}/dE_{\nu}$, is calculated by (e.g.,~\cite{Murase:2015xka})

%===================================
\begin{equation}
E_\nu^2\Phi_\nu(E_\nu)=\frac{c}{4 \pi}
\int_0^{z_{\rm max}}\frac{dz}{1+z}\left|\frac{dt}{dz}\right| 
\left[\varepsilon_\nu^2\frac{d\dot{N}_{\nu}}{d\varepsilon_\nu}(\varepsilon_\nu)\right]n_0\psi(z), 
\label{eq:general_t}
\end{equation}
%===================================
%%===================================
%\begin{equation}
%\frac{d J_{\nu}}{dE_{\nu}}(E_{\nu}) = \frac{c n_0}{4 \pi} 
%\int_0^{\rm z_{\rm max}} 
%dz_s \psi(z) (1 + z) \left| \frac{dt}{dz} \right| 
%\frac{d^2 N_{\nu}}{dt^s dE_{\nu}^s}(E_{\nu}^s, z_s), 
%\label{eq:general_t}
%\end{equation}
%%===================================

where $d\dot{N}_{\nu}/d\varepsilon_\nu$ is the neutrino spectrum per source,
which is calculated in the previous subsection, and $E_\nu=\varepsilon_\nu/(1+z)\approx\Gamma\varepsilon'_\nu/(1+z)$.
The comoving number density of UHECR sources is represented by $n_0\psi(z)$ with the local source density
at $z=0$, $n_0$, and its cosmological evolution factor $\psi(z)$.
For transient sources such as GRBs, $n_0$ is effectively given by $n_0=\rho_0\Delta T$, where $\rho_0$
and $\Delta T$ are the rate density and the duration of neutrino emission at the sources. 

The evolution factor $\psi(z)$ is parameterized as $(1 + z)^m$ such that the parameter $m$ represents
the ``scale'' of the cosmological evolution that is often used in the literature.
In this work, the source evolution is assumed to be compatible with the star formation rate,
which is consistent with the constraints on cosmogenic neutrinos
with extremely-high energy (EHE) analysis by IceCube~\cite{Aartsen:2016ngq}.
Following Refs.~\cite{Kotera:2010yn,Yoshida:2012gf}, we parameterize $\psi(z)$ as

%===================================
\begin{eqnarray}
\psi(z) \propto \left\{ 
\begin{array}{ll}
(1 + z)^{3.4} & ( 0 \leq z \leq 1 ) \\
{\rm constant} & (1 \leq z \leq 4)
\end{array}
\right. .
\label{eq:sdf}
\end{eqnarray}
%===================================

Based on Eq.~(\ref{eq:calorimetric}), $\varepsilon_i^2(d\dot{N}_{\rm CR}/d\varepsilon_i)$
can essentially be regarded as the luminosity of injected cosmic rays.
Then, $n_0\varepsilon_i^2(d\dot{N}_{\rm CR}/d\varepsilon_i)$ corresponds
to the UHECR luminosity density that is known to be
$E_i(dQ_{\rm CR}/dE_i)\approx{10}^{43.8}~{\rm erg}~{\rm Mpc}^{-3}~{\rm yr}^{-1}$~\cite{Katz:2013ooa,Murase:2018utn}.    
The diffuse neutrino flux measurements suggest that the energy generation rate density of neutrinos is comparable,
$E_\nu (dQ_{\nu}/dE_\nu)\approx{10}^{43.3}~{\rm erg}~{\rm Mpc}^{-3}~{\rm yr}^{-1}$~\cite{Murase:2018utn}. 
Both the UHECR and the neutrino diffuse fluxes scale as
$\propto n_0 \varepsilon_{i0} K_{\rm CR}\sim n_0\xi_{\rm CR}L_\gamma \approx n_0\xi_{\rm CR}L'_\gamma \Gamma^2$.
It is convenient to introduce the {\it boosted} source number density defined as

%===================================
\begin{eqnarray}
\mathcal{N}_\Gamma &\equiv& n_0\xi_{\rm CR}\Gamma^2\nonumber\\
&=& \rho_0\Delta T\xi_{\rm CR}\Gamma^2.
\label{eq:boosted_density}
\end{eqnarray}
%===================================

The UHECR and neutrino intensities are then proportional to
$\mathcal{N}_\Gamma$ for a given comoving photon luminosity $L'_\gamma$.
It should be noted that $Q_{\rm CR}(>10~{\rm PeV})=L'_\gamma {\mathcal N}_\Gamma=\xi_{\rm CR}L_\gamma n_0$. 
The full description of $\Phi_\nu$ with the present analytical formulation is given in
Appendix A.

%==========================================
\begin{figure}[t]
  \begin{center}
  \includegraphics[width=0.45\textwidth]{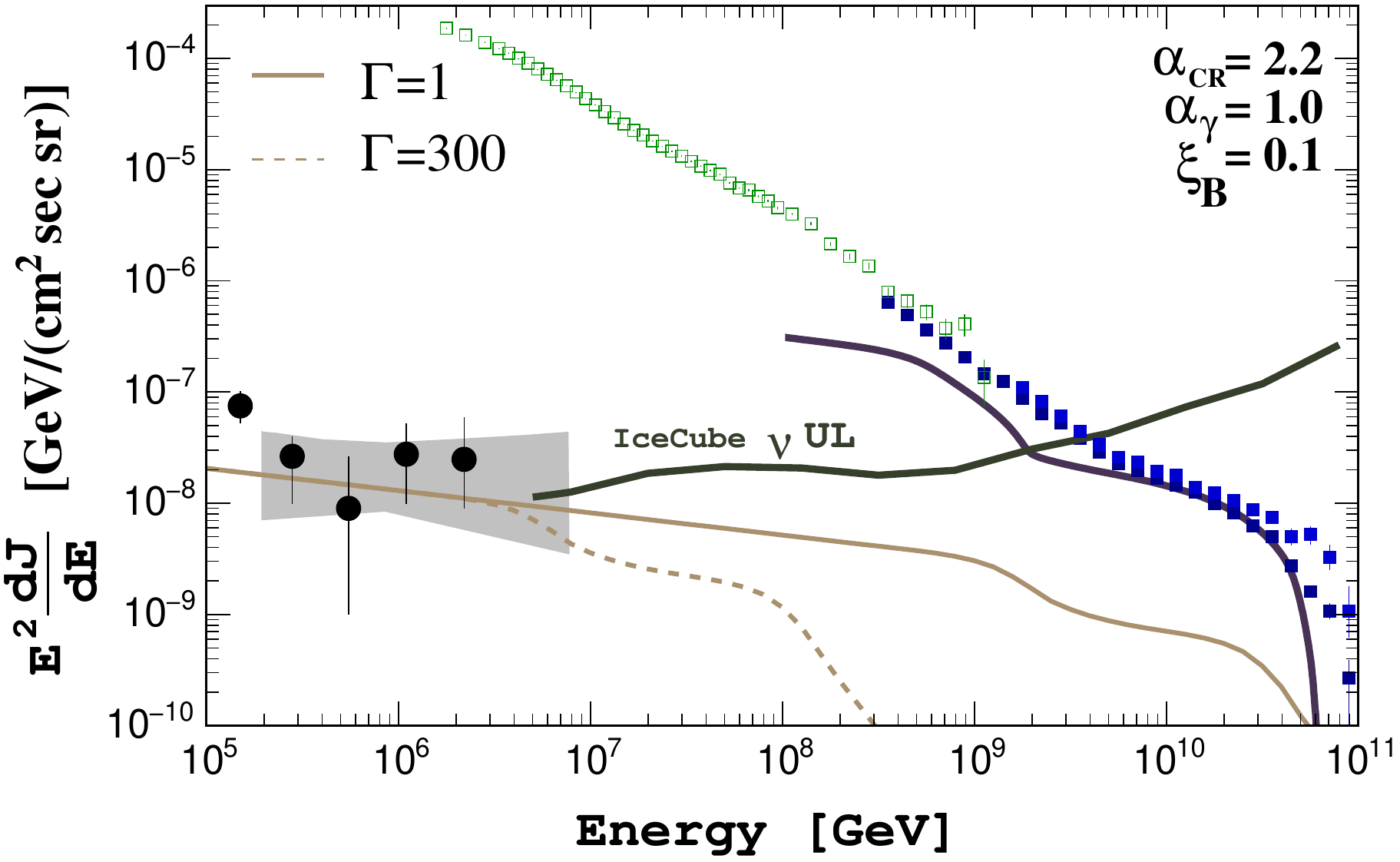}
  \caption{An example of the UHECR nucleon and the all-flavor-sum neutrino fluxes from UHECR sources
    calculated using the presented analysis. The case of $\alpha_{\rm CR}=2.2$, $\alpha_\gamma=1.0$, $\xi_{\rm B}=0.1$ is shown,
    assuming star formation rate-like evolution. The comoving $L'_\gamma$ is set to $4.5\times 10^{46}$ erg/s
    and the boosted source number density $\mathcal{N}_\Gamma$ (Eq.~\ref{eq:boosted_density}) is
    $1\times 10^{-9}$ Mpc$^{-3}$. The optical depth $\tau_{p\gamma0}$ is 0.30 in this example,
    which gives a magnetic field of $B'= 0.91\Gamma^2$~G with $\xi_{\rm B}=0.1$.
    The black points represent the IceCube neutrino measurements~\cite{Aartsen:2015zva} and
    the shaded region represents the flux space, consistent with the IceCube diffuse $\nu_\mu$ data~\cite{Aartsen:2016xlq}.
    The solid curve labeled as (IceCube $\nu$ UL) is the differential EHE bound for IceCube~\cite{Aartsen:2018vtx}.
    The cosmic ray data measured by IceTop~\cite{Aartsen:2013wda}, PAO~\cite{Fenu:2017hlc}
    and TA~\cite{AbuZayyad:2012ru} are also displayed.}
  \label{fig:spectrum_example_1_0}
  \end{center}
\end{figure}
%==========================================

Fig.~\ref{fig:spectrum_example_1_0} shows an example of the UHECR and neutrino fluxes
derived using the presented generic model. 
This realization of the fluxes displayed in Fig.~\ref{fig:spectrum_example_1_0} is associated with
a scenario that is consistent with the UHECR and IceCube data.
We discuss the allowed parameter space in the next section.
It should be noted that the $1/\Gamma$ dependence of the neutrino cut-off energy due to
the pion/muon synchrotron losses is also observed in this plot.

The spectrum of the UHECR protons after their propagation  through intergalactic space to
reach the Earth is calculated using a similar analytical technique.
The details will be described in Appendix B.

%
%==================================================================================
%==================================================================================
\subsection{Observational constraints}
%==================================================================================
%==================================================================================

The predicted neutrino and UHECR spectra must be consistent with their observations.
Qualitatively, the UHECR energy budget constrains the product of $L'_\gamma$ and ${\mathcal N}_\Gamma$,
whereas the neutrino energy budget determines the product of $\tau_{p\gamma0}$ and $L'_\gamma{\mathcal N}_\Gamma$.
To quantify this consistency, we introduce the following criteria in the present study.

%==================================================================================
\begin{enumerate}
  \renewcommand{\theenumi}{(\Roman{enum})}

\item[(a)] The integrated UHECR proton flux above 10 EeV,
  $\int_{10 {\rm EeV}} dE_i dJ_{\rm CR}/dE_i$, is less than the measurement by Auger,
  $8.5\times 10^{-19}$ /cm$^2$/s/sr~\cite{Fenu:2017hlc}\label{cond:Auger}.
  Considering the uncertainties associated with the UHECR mass composition,
  we only request these bolometric requirements of the UHECR flux to be conservative
  with respect to imposing bounds on the relevant parameter space. The results for the required UHECR energy generation rate
  density are consistent with those obtained based on detailed numerical simulations considering the uncertainties. 

%\km{Apparently there is some difference with. This condition is used for constraints on the energy budget, i.e., $L'_\gamma N_\Gamma$
%My sophisticated analysis based on the propagation code indicate $EdQ/dE\approx0.6-1.2\times{10}^{44}(\alpha_{\rm CR}-1)\times{10}^{44}~{\rm erg}~{\rm Mpc}^{-3}~{\rm yr}^{-1}$ at $10^{19.5}$~eV, which leads to $Q\approx(1.8-3.6)\times{10}^{45}~{\rm erg}~{\rm Mpc}^{-3}~{\rm yr}^{-1}$ for $\alpha_{\rm CR}=2.2$ and $Q\approx(2.9-5.8)\times{10}^{45}~{\rm erg}~{\rm Mpc}^{-3}~{\rm yr}^{-1}$ for $\alpha_{\rm CR}=2.3$. 
%In the case of pure silicon, the propagation code indicates $EdQ/dE\approx0.45\times{10}^{44}(\alpha_{\rm CR}-1)\times{10}^{44}~{\rm erg}~{\rm Mpc}^{-3}~{\rm yr}^{-1}$ at $10^{19.5}$~eV, $Q\approx2.2\times{10}^{45}~{\rm erg}~{\rm Mpc}^{-3}~{\rm yr}^{-1}$ for $\alpha_{\rm CR}=2.3$.}

\item[(b)] The neutrino flux intensity at 100 TeV and the spectral power law index are
  within the 99 \% C.L. range obtained by the diffuse $\nu_\mu$ data
  measured by IceCube~\cite{Aartsen:2016xlq}~\label{cond:diffuseNuMu}.

\item[(c)] The all-flavor-sum neutrino flux at 100 PeV is less than $2\times 10^{-8}$ GeV/cm$^2$/s/sr,
  the limit obtained by the IceCube EHE analysis~\cite{Aartsen:2018vtx}.
  
\item[(d)] The neutrino flux at 6 PeV is above $2\times 10^{-9}$ GeV/cm$^2$/s/sr,
  determined by the 6 PeV energy neutrino detection by IceCube~\cite{Aartsen:2018vtx}\label{cond:6PeV}.
  
\end{enumerate}
%==================================================================================

%==================================================================================
%==================================================================================
%==================================================================================
\section{Constraints on UHECR and Neutrino Emitters \label{sec:results}}
%==================================================================================
%==================================================================================
%==================================================================================

Four constraints from UHECR acceleration (Eqs.~\ref{eq:hillas} and \ref{eq:sync_condition}),
UHECR escape (Eqs.~\ref{eq:esc_condition} and \ref{eq:calorimetric})
based on the physics of photo-meson production, and diffuse UHECR and neutrino flux measurements,
allow us to constrain generic unification models for photohadronic neutrinos. 
We present the results in the following.

%==================================================================================
%==================================================================================
\subsection{Cases of fiducial neutrino spectra \label{subsec:parameter_general}}
%==================================================================================
%==================================================================================

%==========================================
\begin{figure*}
\begin{center}
\includegraphics[width=0.45\textwidth]{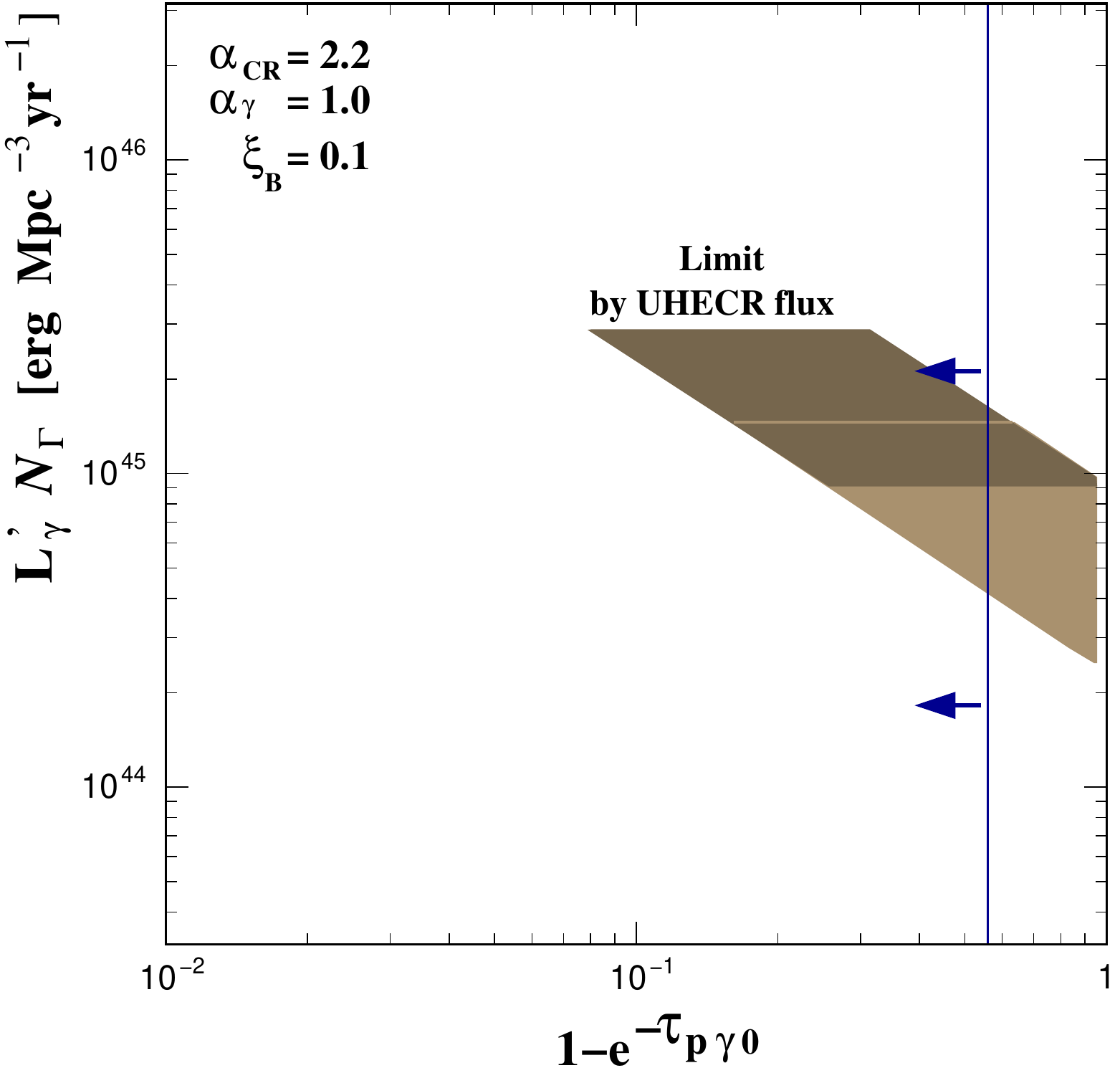}
\includegraphics[width=0.45\textwidth]{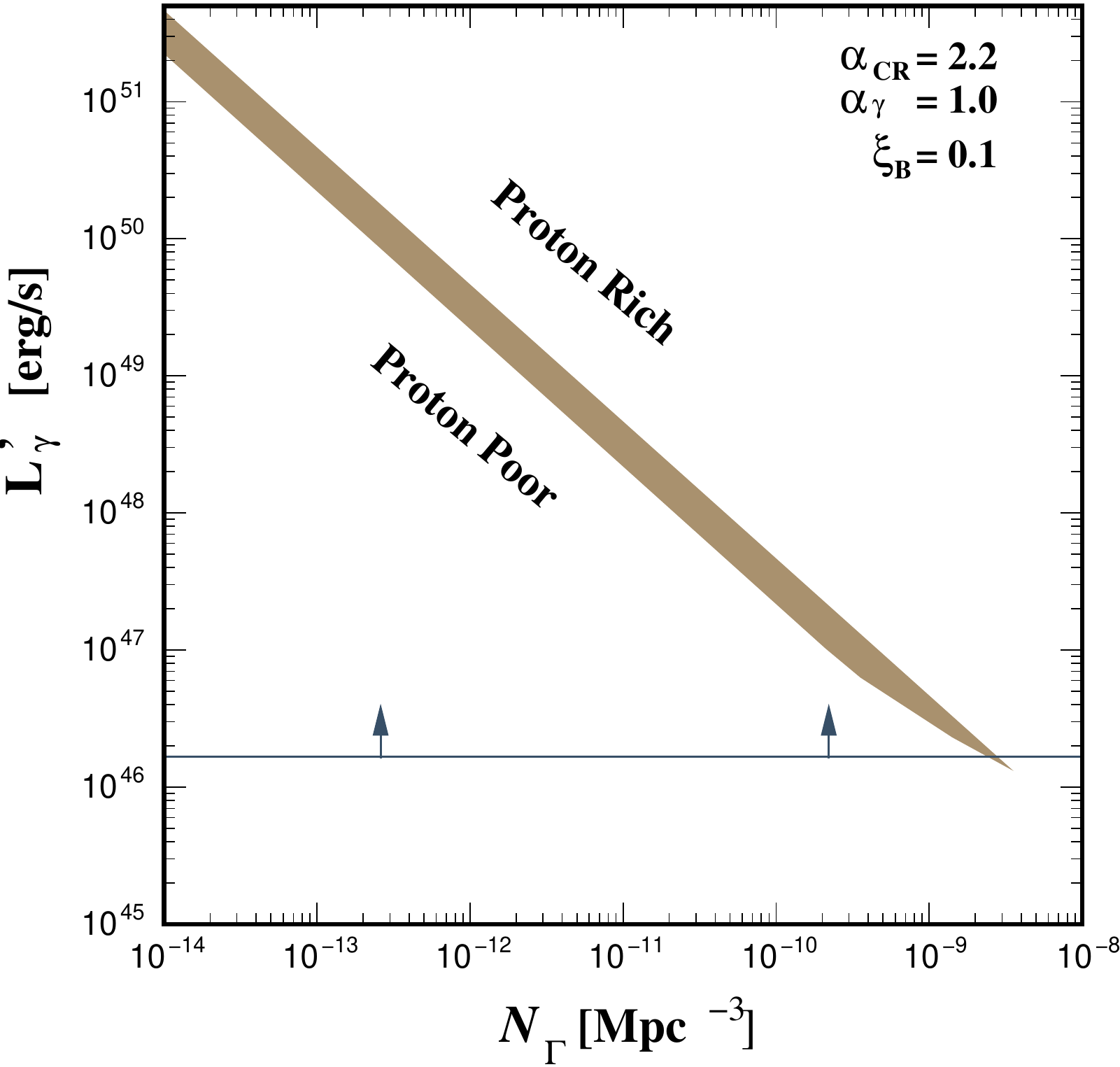}
\caption{(Left) The allowed region in the parameter space of luminosity per unit volume, $L'_\gamma\mathcal{N}_\Gamma$,
  and damping factor $\displaystyle{1-e^{-\tau_{p\gamma0}}}$.
  The parameters inside the shaded region satisfy the observational consistency criteria of
  conditions~(a)$\sim$~(d) described in the text, and the UHECR condition of Eq.~(\ref{eq:sync_condition}).
  The cases of $\alpha_{\rm CR}=2.2$ and $\gamma=1.0$ are shown. We find no $\Gamma$ dependence on these constraints.
  The horizontal belt represented by the darker shade shows the systematics of the UHECR energetics
 that originate from the uncertainties on the mass composition and Galactic to the extragalactic transition of
  UHECRs~\cite{Murase:2018utn}. The vertical line represents the bound on $\tau_{p\gamma0}$
  by the UHECR escape condition, Eq.~(\ref{eq:esc_condition}) when $\xi_{\rm B}=0.1$.
  The maximal bound of $L'_\gamma\mathcal{N}_\Gamma$ is determined by the condition whereby
  the proton flux from sources should not exceed the measured flux of UHECRs.
  The lower bound is driven by the intensity of neutrinos measured by IceCube.
  (Right) The allowed region on the plane of the source luminosity $L'_\gamma$ and
  the boosted source density $\mathcal{N}_\Gamma$.
  The parameters inside the shaded region statisfy the observational consistency criteria
  as shown in the left plot. The horizontal line represents the condition of $t_p^{\rm acc}\leq t_p^{\rm dyn}$, Eq.~(\ref{eq:hillas}).
}
\label{fig:constraints_when_alpha2_2}
\end{center}
\end{figure*}
%==========================================

%==========================================
\begin{figure*}
\begin{center}
\includegraphics[width=0.45\textwidth]{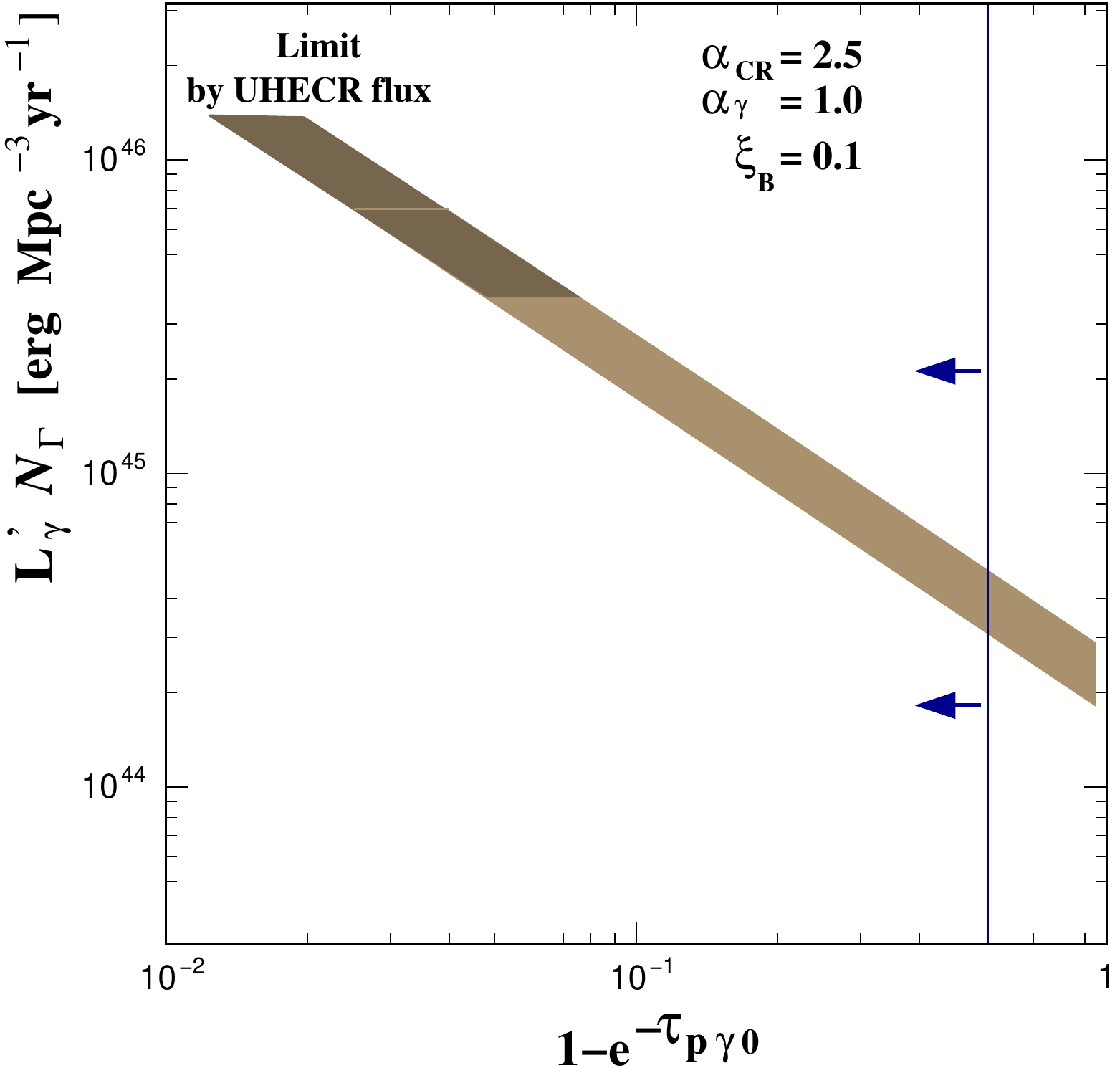}
\includegraphics[width=0.45\textwidth]{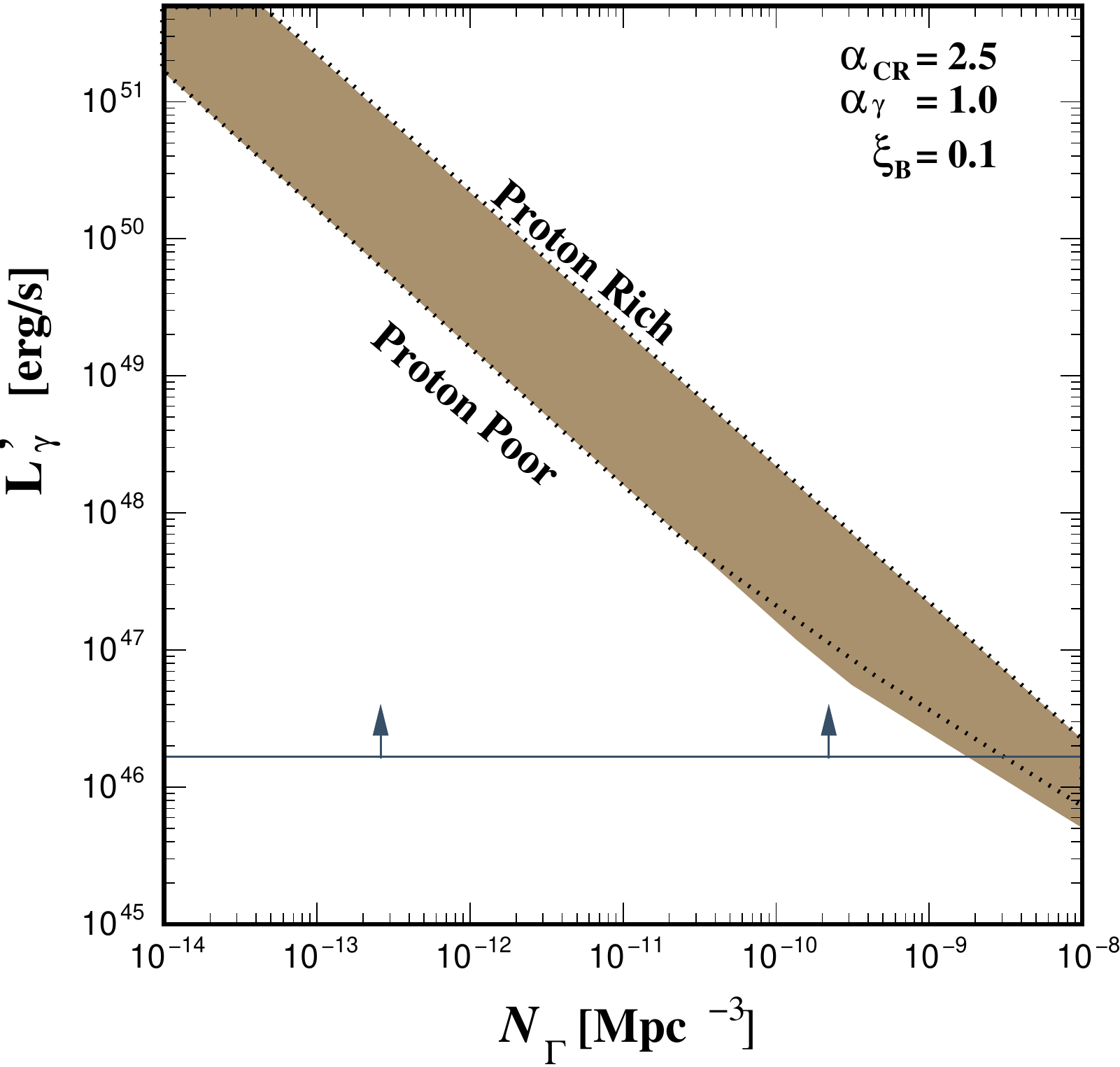}
\caption{Same as Fig.~\ref{fig:constraints_when_alpha2_2} but for $\alpha_{\rm CR}=2.5$.
  The constraints on $L'_\gamma$--$\mathcal{N}_\Gamma$ (right) has a small dependence on $\Gamma$ when $\Gamma\gg 1$.
  The region specified by the dashed line corresponds to the allowed space for $\Gamma=100$.
}
\label{fig:constraints_lumonisoty_when_alpha2_5}
\end{center}
\end{figure*}
%==========================================

The left plot of Fig.~\ref{fig:constraints_when_alpha2_2}
displays the luminosity and the optical depth constraints for the spectral power law index of UHECRs
$\alpha_{\rm CR}=2.2$ and that of the target photons $\alpha_\gamma=1.0$.
Given that the neutrino spectrum follows $\propto E_\nu^{-(\alpha_{\rm CR}-\alpha_\gamma+1)}\sim E_\nu^{-2.2}$
(see Eq.~(\ref{eq:onsource_final})), they represent the case of neutrino spectra with $\alpha_\nu>2$,
which is close to the index suggested by IceCube observations. 
The optical depth $\tau_{p\gamma0}\gtrsim 0.1$ is required because
the IceCube neutrino energy flux is compatible with the UHECR flux.
As seen in Fig.~\ref{fig:spectrum_example_1_0}, the margins for the neutrino fluxes to be consistent
with both the neutrino and UHECR observations are small
when the primary UHECR spectrum is as hard as $\alpha_{\rm CR}\lesssim 2.2$.
Since $L'_\gamma\mathcal{N}_\Gamma\propto n_0K_{\rm CR}$, the range of the luminosity per unit volume,
$L'_\gamma\mathcal{N}_\Gamma$, is bounded by the UHECR flux and the IceCube neutrino flux
connected by the optical depth $\tau_{p\gamma0}$.
This is an expanded way of presenting the bounds leading to the
frequently referenced Waxman-Bahcall limit~\cite{Waxman:1998yy}. 
The tight constraint is also consistent with the results in Ref.~\cite{Yoshida:2014uka}.
The resultant range of the source luminosity per unit volume
is $\sim (3-15)\times 10^{44}~{\rm erg}~{\rm Mpc}^{-3}~{\rm yr}^{-1}$.
This is comparable with the integrated UHECR luminosity per unit volume at $z=0$ above $10^{18}$~eV~\cite{Decerprit:2011qe}.

However, the UHECR escape condition, Eq.~(\ref{eq:esc_condition}),
prevents large optical depths unless the magnetic field is weaker than expected from the equipartition condition
$\xi_{\rm B}=1$. The bound of $\tau_{p\gamma0}\lesssim 0.06$ derived by Eq.~(\ref{eq:esc_condition})
with the equipartition condition $\xi_{\rm B}=1$ is obviously inconsistent with the shaded region
in the left plot of Fig.~\ref{fig:constraints_when_alpha2_2}.
Relaxation of the criteria for proton synchrotron cooling
by setting $\xi_{\rm B}\sim 0.1$ can open an allowed space of the parameters,
$L'_\gamma\mathcal{N}_\Gamma$ and the optical depth $\tau_{p\gamma0}$.
We found that the cases of even harder UHECR source spectrum, {\it i.e.},
$\alpha_{\rm CR}\lesssim 2.1$ is nearly excluded for a reasonable range of the magnetic field strengths
that are expected due to $\xi_{\rm B}\gtrsim 0.1$. Given that the upper bound of $\tau_{p\gamma 0}$ required by the UHECR escape condition
scales as $1/\varepsilon_i^{\rm max}$, ({\it c.f.} Eq.~(\ref{eq:esc_condition})),
setting $\varepsilon_i^{\rm max}\ll 10^{11}$~GeV relaxes these constraints.

We also found that the allowed range of optical depths is limited,
yielding $0.1\lesssim \tau_{p\gamma0}\lesssim 0.6$ for a given value of $\xi_B\sim0.1$,
and it is even more severely constrained if $\xi_B\gg 0.1$. This is nearly a universal bound
regardless of the UHECR spectral index if $\alpha_{\rm CR}\lesssim 2.3$.

The right plot of Fig.~\ref{fig:constraints_when_alpha2_2} shows the allowed parameter space
on the source luminosity in the plasma rest frame $L'_\gamma$ and the boosted source number density $\mathcal{N}_\Gamma$.
For the requirement of the luminosity condition, Eq.~\ref{eq:hillas},
the unified sources must be relatively rare, $\mathcal{N}_\Gamma\lesssim 10^{-9}~{\rm Mpc}^{-3}$.
This is a well-known consequence of the UHECR energy budget argument.
The minimal value of $L'_\gamma$ in the shaded region is determined by the synchrotron cooling condition,
$t'_{\rm acc}<t'_{\rm syn}$, Eq.~(\ref{eq:sync_condition}),
but the lower bound of $L'_\gamma$ demanding $t'_{\rm acc}<t'_{\rm dyn}$, Eq.~(\ref{eq:hillas}), is more stringent.

We note that these constraints in the plane of luminosity per unit volume and
the optical depth, and the plane of $L'_\gamma$--$\mathcal{N}_\gamma$ are 
nearly independent of the plasma bulk Lorentz factor $\Gamma$.
Thus, they are universal conditions that any class of sources in a unification scheme should satisfy.

The constraints on the source luminosity per unit volume $L'_\gamma\mathcal{N}_\gamma$
can be relaxed for the case of the {\it soft} UHECR (and thus neutrino) spectra.
Fig.~\ref{fig:constraints_lumonisoty_when_alpha2_5} displays an example, $\alpha_{\rm CR}=2.5$. 
Since the margin between UHECR and the neutrino fluxes increases if the UHECR proton spectrum is steeper,
the luminosity per volume can be $\gtrsim 3\times 10^{45}~{\rm erg}~{\rm Mpc}^{-3}~{\rm yr}^{-1}$.
The sources that satisfy this requirement for CRs include galaxies, AGNs, and, galaxy clusters~\cite{Murase:2018utn}.

%==================================================================================
%==================================================================================
\subsection{Cases of hard neutrino spectra \label{subsec:parameter_relativistic}}
%==================================================================================
%==================================================================================

%==========================================
\begin{figure}
\begin{center}
\includegraphics[width=0.45\textwidth]{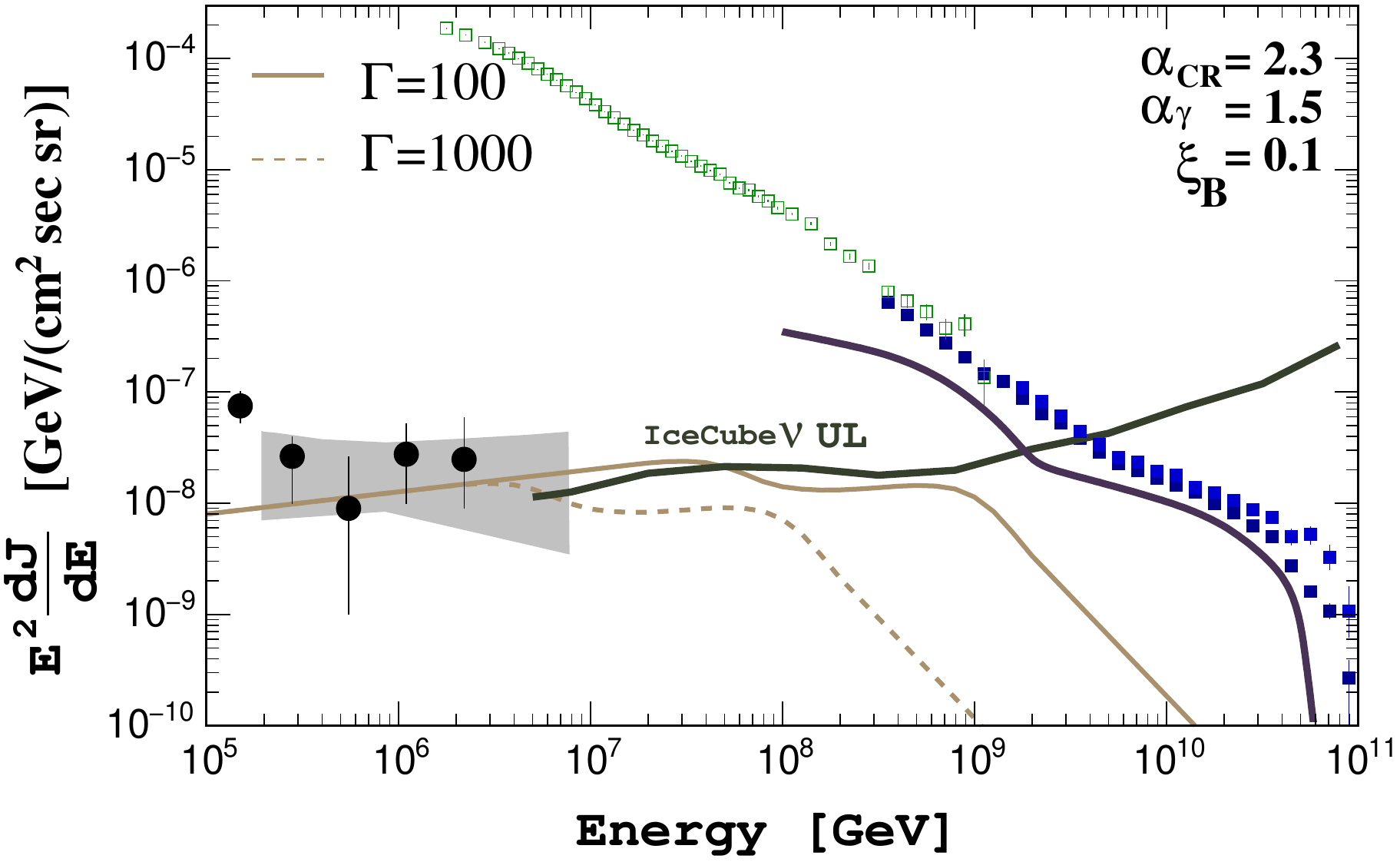}
\caption{An example of scenario for hard neutrino flux. $\alpha_{\rm CR}=2.3$ and $\alpha_\gamma=1.5$.
  The comoving $L'_\gamma$ is set to $5.0\times 10^{48}$ erg/s and the boosted source number density
  $\mathcal{N}_\Gamma$ (Eq.~\ref{eq:boosted_density}) is $1\times 10^{-9}$ Mpc$^{-3}$.
  The optical depth $\tau_{p\gamma0}$ is 0.10 in this particular example,
  we have a magnetic field of $B'= 0.26\Gamma^2$~G with $\xi_{\rm B}=0.1$.}
\label{fig:hard_flux_scenario}
\end{center}
\end{figure}
%==========================================
%==========================================
\begin{figure*}
  \begin{center}
  \includegraphics[width=0.45\textwidth]{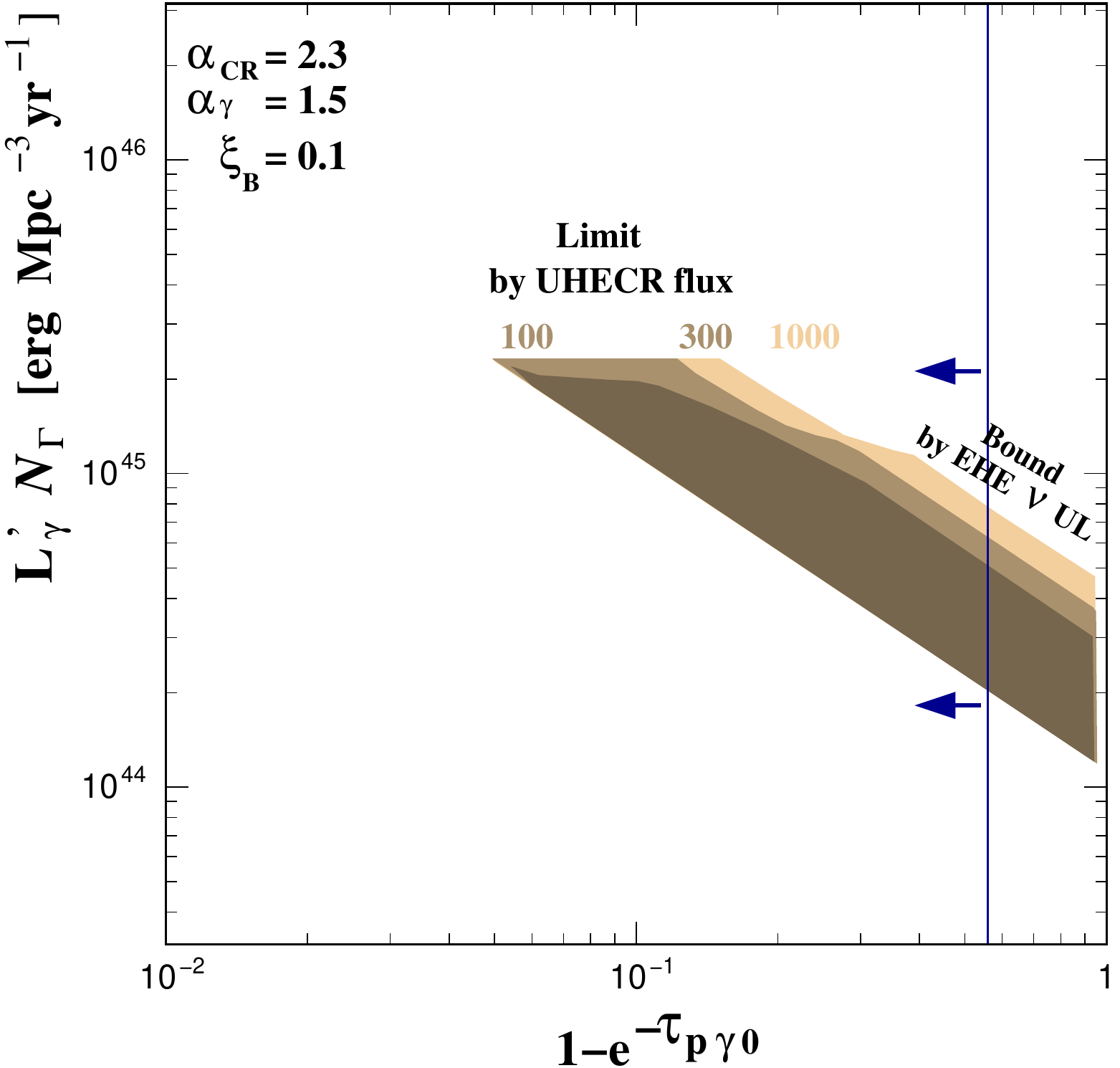}
  \includegraphics[width=0.45\textwidth]{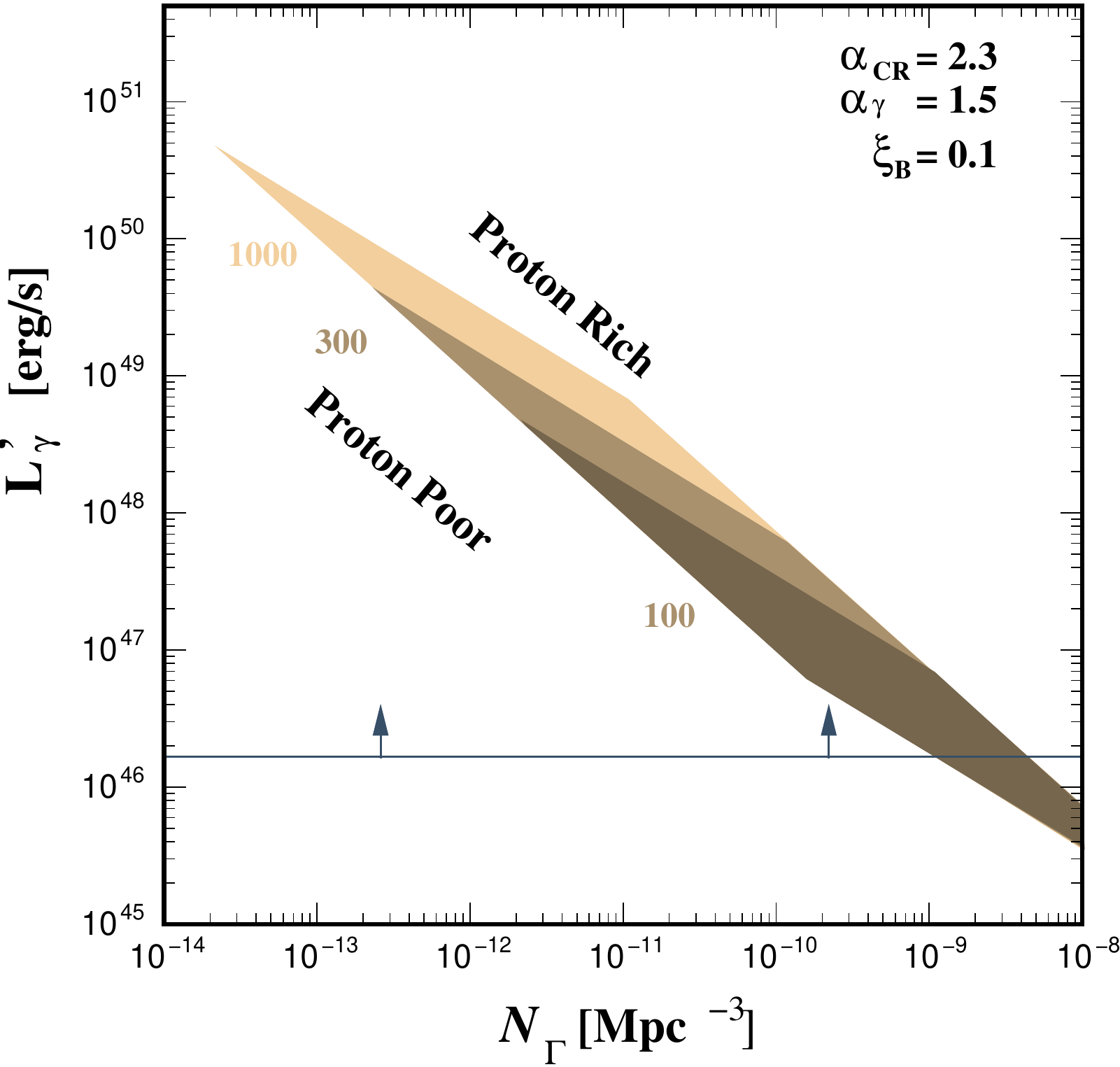}
  \caption{Same as Fig.~\ref{fig:constraints_when_alpha2_2}
    but with $\alpha_{\rm CR}=2.3$ and $\alpha_\gamma=1.5$.
    Only a relativistic plasma flow, $\Gamma\gtrsim 30$, can be consistent with the observation and
    the resultant allowed region has weak dependences on $\Gamma$. In these plots,
    the allowed parameter space for $\Gamma=100$, $\Gamma=300$, and $\Gamma=1000$ are represented using
    different shades.}
  \label{fig:constraints_hard_flux_scenario}
  \end{center}
\end{figure*}
%==========================================

%==========================================
\begin{figure}
\begin{center}
\includegraphics[width=0.45\textwidth]{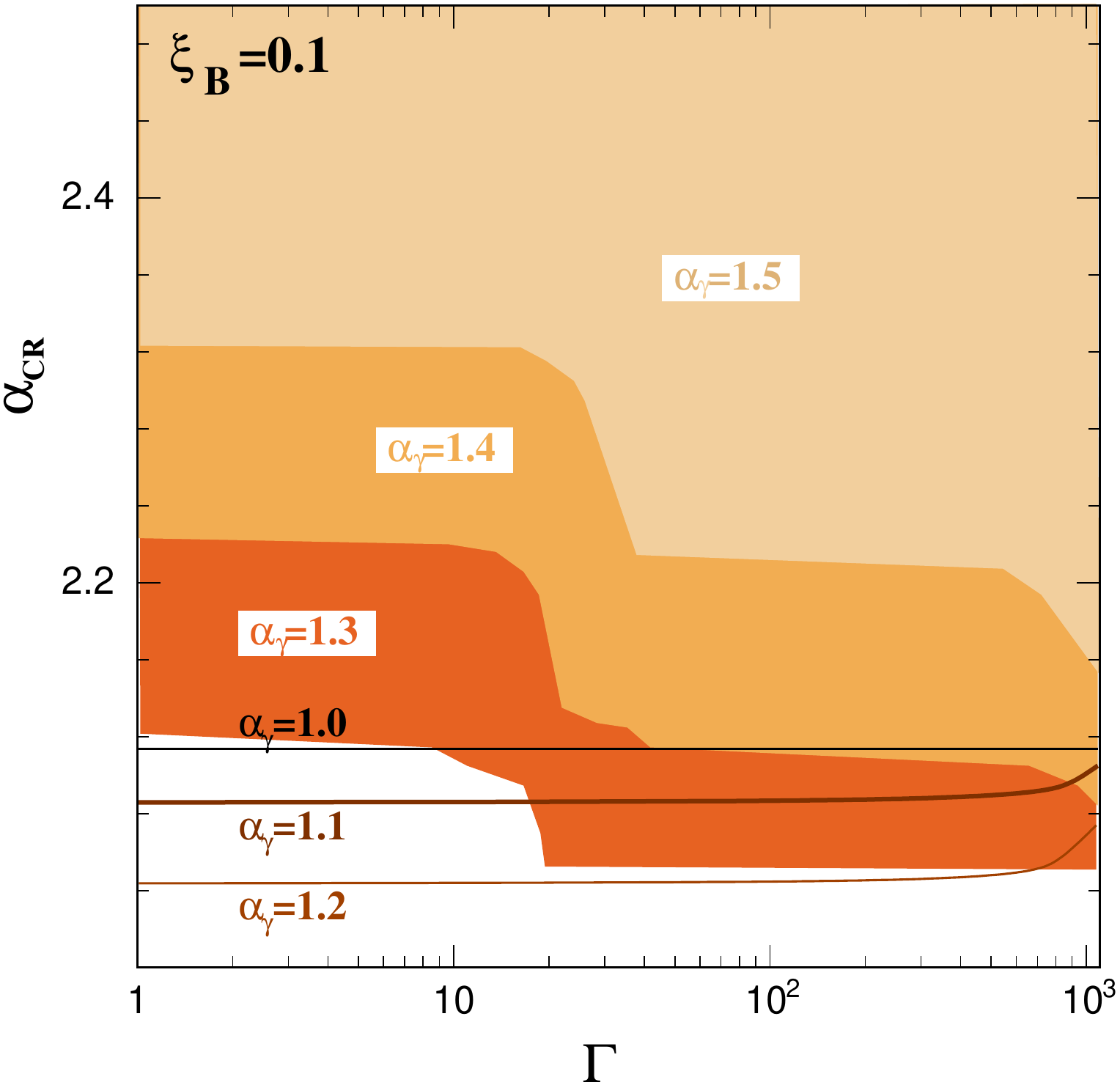}
\caption{The allowed region in the plane of $\alpha_{\rm CR}$ and $\Gamma$.
  The regions for $\alpha_\gamma=1.4$ and $\alpha_\gamma=1.3$ and $\alpha_\gamma=1.5$ are
  represented by different shades. The cases of $\alpha_\gamma=1.0$, 1.1, and 1.2 are represented by the solid curve.
  The region above each of the lines is allowed. $\xi_{\rm B}=0.1$ is assumed.
}
\label{fig:constraints_on_lorentz_factor}
\end{center}
\end{figure}
%==========================================

Although cases of harder UHECR spectra, {\it i.e.}, $\alpha_{\rm CR}\lesssim 2.1$ are nearly eliminated,
a scenario that predicts hard {\it neutrino} spectra with $\alpha_\nu\lesssim2.0$ is more realistic
if the target photon spectrum is softer as $\alpha_\gamma\gtrsim 1.3$. 
It should be noted that the neutrino spectrum follows $\sim E_\nu^{-(\alpha_{\rm CR}-\alpha_\gamma+1)}$ according to Eq.~(\ref{eq:onsource_final}). 
A hard neutrino spectrum like $\sim E_\nu^{-2}$ cannot extend well above 100 PeV and
should attenuate at a point below this value, given that the spectral extension to $\gg~{\rm PeV}$
with a $E_\nu^{-2}$-like power-law flux has been eliminated by the IceCube EHE limit
(see Fig.~3 of Ref.~\cite{Aartsen:2016ngq}).
The spectral fall-off behavior of the neutrino spectrum is naturally expected when strong synchrotron cooling occurs.
As discussed earlier, since the characteristic synchrotron cut-off energy of neutrinos is
$\varepsilon_\nu^{\rm syn}\sim \sqrt{L'_\gamma}/(\Gamma\tau_{p\gamma0})$, a lower energy cut-off via synchrotron cooling
is realized in relativistic plasma flow {\it i.e.}, $\Gamma\gg 1$. 
A scenario of harder neutrino spectra (but softer UHECR spectra) is, therefore,
a natural consequence of the unified UHECR/neutrino model for ultra-relativistic sources.

An example of the ultra-relativistic scenario is shown in Fig.~\ref{fig:hard_flux_scenario}.
The hard neutrino spectrum $\propto E_\nu^{-(\alpha_{\rm CR}-\alpha_\gamma+1)}\sim E_\nu^{-1.8}$ falls off
at $\sim 500~(50)~{\rm PeV}$ for sources with $\Gamma=100~(1000)$.
These spectra are consistent with the IceCube EHE limit~\cite{Aartsen:2018vtx}
based on the null detection of $\gtrsim 10~{\rm PeV}$ neutrinos.
They represent a scenario of ultra-relativistic sources with unified UHECR and neutrino emission.

Since the cut-off energy of the neutrino spectrum depends explicitly on $\Gamma$ for a given optical depth,
the constraints on $L'_\gamma, \mathcal{N}_\gamma$, and $\tau_{p\gamma0}$ exhibits a weak dependence on $\Gamma$
in the case of extremely relativistic sources that yield hard neutrino fluxes.
Fig.~\ref{fig:constraints_hard_flux_scenario} displays the allowed region of parameter spaces
in the hard neutrino spectrum. Since $E_\nu^{\rm syn}\propto \sqrt{L'_\gamma}/(\Gamma\tau_{p\gamma0})$,
a lower $\Gamma$ excludes super-luminous sources, because the neutrino intensity at $\gg~{\rm PeV}$
would overshoot the IceCube EHE limit.

Given that the spectral indexes $\alpha_{\rm CR}$ and $\alpha_\gamma$ characterize the emission environments,
it is important to understand their allowed space in the unified source model.
Fig.~\ref{fig:constraints_on_lorentz_factor} shows the constraints in the plane of $\alpha_{\rm CR}$ and $\Gamma$
for various values of the photon spectral power-law index $\alpha_\gamma$. 
The rapid fall-off structures observed at $\Gamma\sim 20$ result from the spectral cut-off due to synchrotron cooling.
A higher $\Gamma$ facilitates larger parameter spaces of $\alpha_{\rm CR}$ and $\alpha_\gamma$
as it avoids the EHE neutrino limit.
The figure also indicates that extremely relativistic cases, $\Gamma\sim 10^3$,
would further extend the allowed parameter space. This is because strong synchrotron cooling
softens a fairly hard spectrum of neutrinos, which would otherwise be inconsistent with the IceCube observation.

\begin{figure}
\begin{center}
\includegraphics[width=0.45\textwidth]{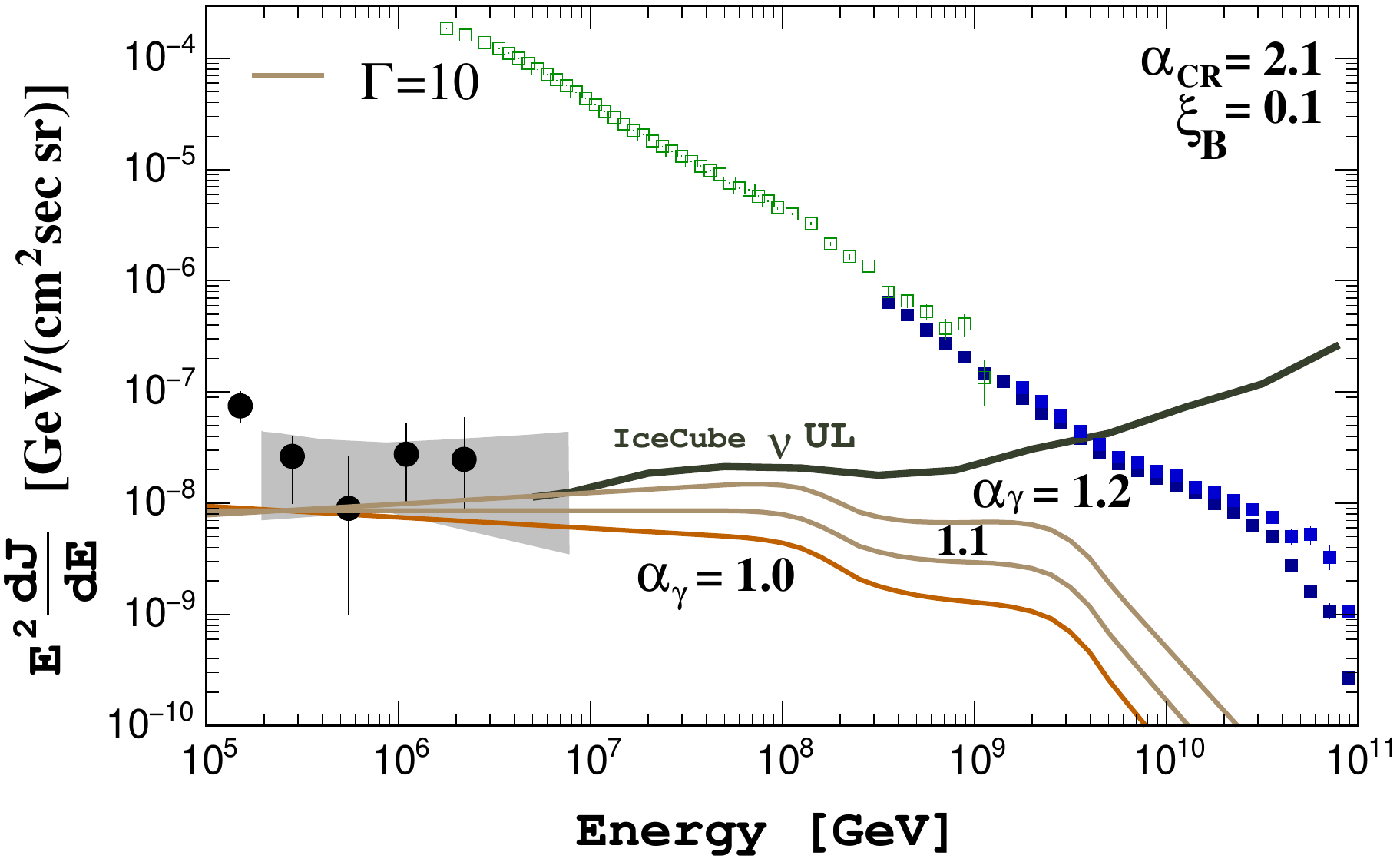}
\caption{An example of neutrino energy spectra from the hard UHECR flux with $\alpha_{\rm CR}=2.1$.
  In this example, $L'_\gamma$, $N_\Gamma$, and $\tau_{p\gamma0}$ are  $8.4\times 10^{46}~{\rm erg}~{\rm s}^{-1}$,
  $3.0\times 10^{-10}$~Mpc$^{-3}$, and 0.42, respectively, which give a magnetic field of
  $B'= 0.91\Gamma^2$~G with $\xi_{\rm B}=0.1$.
  The case of $\alpha_\gamma=1.0$ is not consistent with the IceCube diffuse $\nu_\mu$ analysis~\cite{Aartsen:2016xlq}.}
\label{fig:hard_UHECR_flux_scenario}
\end{center}
\end{figure}

Fig.~\ref{fig:constraints_on_lorentz_factor} also indicates that harder UHECR proton emission
$\alpha_{\rm CR}\lesssim 2.1$ is nearly excluded, as discussed earlier.
This bound depends on the photon spectral index $\alpha_\gamma$ in a non-trivial way.
The situation is illustrated in Fig.~\ref{fig:hard_UHECR_flux_scenario}.
The cases of $\alpha_\gamma=1.1$ and $1.2$ are allowed but $\alpha_\gamma=1.0$ is inconsistent
because it is too soft to be allowed in the diffuse $\nu_\mu$ observations (condition~(b) described in Sec.~\ref{sec:condition}).
The IceCube data favors a harder spectrum when we assume a lower side of the allowed intensity region
$\displaystyle{E_\nu^2/dJ_{\nu_e+\nu_\mu+\nu_\tau}/dE_\nu\sim 1\times 10^8~{\rm GeV}~{\rm cm}^{-2}~{\rm s}^{-1}~{\rm sr}^{-1}}$
which is the only possibility that allows for consistency with the UHECR flux.

%==========================================
%==========================================

%==================================================================================
%==================================================================================
\subsection{Cases of UHECR nuclei}
%==================================================================================
%==================================================================================

The recent observations by Auger indicate that UHECRs are likely to have a mixed composition
that is dominated by intermediate to heavy nuclei at the highest energies. 
This adds important conditions to the possible classes of the sources.
That is, we require that nuclei with $A>1$ and $Z>1$ are accelerated and survive.
The maximum proton energy can be lower, but the survival conditions constrain
the source environments more strongly as investigated for GRBs~\cite{Murase:2008mr,Horiuchi:2011zz}
and AGNs~\cite{Murase:2011cy,Pe'er:2009rc}.   

The luminosity requirement (Eq.~\ref{eq:hillas}) is significantly relaxed in the cases of heavy nuclei.
The condition by $t_{\rm acc}<t_{\rm syn}$ is similarly imposed via Eq.~(\ref{eq:sync_condition}). 
The new requirements originate from the photodisintegration of nuclei.
As in the proton case, we focus on the situation wherein the system is
effectively optically thin to photodisintegration and photo-meson production,
in which $t'_{\rm acc}<t'_{\rm dis}$ is automatically satisfied. In this case, $t'_{\rm dis}$ is the photodisintegration energy loss time.

After the nuclei are accelerated, they must survive photodisintegration while they leave the sources.
The survival condition is more severe~\cite{Murase:2008mr,Murase:2010gj}.
The photodisintegration cross-section is larger than that for photo-meson production, which gives the optical depth:

%===================================
\begin{equation}
\tau_{A\gamma}(\varepsilon_i)\approx\frac{2}{1+\alpha_\gamma}\frac{L'_{\gamma0}}{4\pi R\Gamma c \varepsilon'_{\gamma0}}
\left(\int ds \frac{\sigma_{A\gamma}(s)}{s-m_A^2}\right)
{\left(\frac{\varepsilon_i}{\tilde{\varepsilon}^{\rm GDR}_{i0}}\right)}^{\alpha_\gamma-1}
\label{eq:optical_depth_nuclei}
\end{equation}
%===================================

where $\tilde{\varepsilon}^{\rm GDR}_{i0}$ is introduced as
%===================================
\begin{eqnarray}
\tilde{\varepsilon}^{\rm GDR}_{i0}&=&\frac{s_{\rm GDR}-m_A^2}{4}\frac{\Gamma}{\varepsilon'_{\gamma0}}\nonumber\\
&=&\frac{s_{\rm GDR}-m_A^2}{s_{\Delta}-m_p^2}{\tilde{\varepsilon}^{\Delta}_{p0}},
\end{eqnarray}
%===================================

where $s_{\rm GDR}=m_A^2+2 m_A \bar{\varepsilon}_{\rm GDR}$ is the Mandelstam variable at the giant dipole resonance,
where $\bar{\varepsilon}_{\rm GDR}\approx42.65A^{-0.21}$~MeV is the resonance energy.   
The photodisintegration process is dominated by the giant dipole resonance.  
This relates $\tau_{A\gamma}$ to $\tau_{p\gamma}$ as

%===================================
\begin{equation}
\tau_{p\gamma0}\approx\tau_{A\gamma}(\varepsilon_i^{\rm max})\frac{\int ds \frac{\sigma_{p\gamma}(s)}{s-m_p^2}}
{\int ds \frac{\sigma_{A\gamma}(s)}{s-m_A^2}}
{\left[\left(\frac{s_{\rm GDR}-m_A^2}{s_\Delta-m_p^2}\right)
\left(\frac{\tilde{\varepsilon}^{\Delta}_{p0}}{\varepsilon_{i}^{\rm max}}\right)\right]}^{\alpha_\gamma-1}
\label{eq:relation_between_pgamma_dis}
\end{equation}
%===================================

The importance of this relationship was highlighted in Ref.~\cite{Murase:2008mr}
(see also Eq.~6 of Ref.~\cite{Murase:2010gj}). 
The survival condition is imposed by $t'_{\rm dyn}<t'_{\rm dis}$, which leads to 

%===================================
\begin{equation}
\tau_{A\gamma}(\varepsilon_i^{\rm max})\lesssim A, 
\end{equation}
%===================================

which is analogous to Eq.~(\ref{eq:calorimetric}). We get

%====================================
\begin{equation}
\tau_{p\gamma0}\lesssim A\frac{\int ds \frac{\sigma_{p\gamma}(s)}{s-m_p^2}}
{\int ds \frac{\sigma_{A\gamma}(s)}{s-m_A^2}}
{\left[\left(\frac{s_{\rm GDR}-m_A^2}{s_\Delta-m_p^2}\right)
\left(\frac{\tilde{\varepsilon}^{\Delta}_{p0}}{\varepsilon_{i}^{\rm max}}\right)\right]}^{\alpha_\gamma-1}.
\label{eq:suvival_condition}
\end{equation}
%====================================

In particular, for $\alpha_\gamma=1.0$, 
this leads to $\tau_{p\gamma}\sim\tau_{p\gamma0}\lesssim0.4~{(A/56)}^{0.79}$, which is equivalent to Eq.~10 of Ref.~\cite{Murase:2010gj}. (
Note that the value itself can be enhanced by the quasideutron process, baryon resonances and photofragmentation.). 
We require this survival condition in addition to Eqs.~(\ref{eq:hillas}),
(\ref{eq:sync_condition}) and (\ref{eq:esc_condition}). 
It should be noted that this constraint is stronger for $\alpha_\gamma>1$.

The aforementioned requirements of the sources are applied independently of the details of the UHECR composition. 
However, the constraints from the diffuse UHECR and neutrino fluxes depend on the composition. 
Even if UHECRs are dominated by nuclei, the lower-energy cosmic rays that are responsible for IceCube neutrinos may be proton dominated, 
in which the diffuse constraints remain unchanged from those obtained in the previous subsections. 
However, if the cosmic rays are dominated by heavy nuclei even at lower energies, 
the constraints are modified. We hereby consider such cases. 
The astrophysical neutrino flux from the UHECR nuclei can be approximately described using a treatment
similar to the case of proton-dominated UHECRs, 
if the UHECR sources are effectively transparent to the photodisintegration process. 
The neutrino flux due to photomeson production via secondary nucleons and primary nuclei is given by
%===================================
\begin{eqnarray}
E_\nu^2\frac{d J_\nu}{d E_\nu}&\approx&\frac{3}{8}[1-{(1-\kappa_{p\gamma})}^{\tau_{p\gamma}}][1-{(1-\kappa_{\rm dis})}^{\tau_{A\gamma}}]E_i^2\frac{d J_{\rm CR}}{d E_i}\nonumber\\
&+&\frac{3}{8}[1-{(1-\kappa_{\rm mes})}^{\tau_{\rm mes}}]
{(1-\kappa_{\rm dis})}^{\tau_{A\gamma}}E_i^2\frac{d J_{\rm CR}}{d E_i}
\end{eqnarray}
%===================================
Note that in the limit of $\kappa_{p\gamma}\ll 1$ and $\kappa_{\rm dis}\ll 1$ keeping $\kappa_{p\gamma}\tau_{p\gamma}<1$ and $\kappa_{A\gamma}\tau_{A\gamma}<1$, we have
\begin{eqnarray}
E_\nu^2\frac{d J_\nu}{d E_\nu}&\approx&\frac{3}{8}\kappa_{p\gamma}\tau_{p\gamma}[E_i/A] \kappa_{\rm dis}\tau_{A\gamma}E_i^2\frac{d J_{\rm CR}}{d E_i}\nonumber\\
&+&\frac{3}{8}\kappa_{\rm mes}\tau_{\rm mes}[E_i]
(1-\kappa_{\rm dis}\tau_{A\gamma})E_i^2\frac{d J_{\rm CR}}{d E_i},
\end{eqnarray}
which is similar to Eq.~(11) of Ref.~\cite{Murase:2010gj}. The first term of the right hand side represents the contribution from secondary nucleons while the second term is for the contribution from the photomeson production on nuclei.

With $\tau_{\rm mes}[E_i]\sim A\tau_{p\gamma}[E_i/A]$ (because of the approximation, $\sigma_{\rm mes}[E_i]\sim A\sigma_{p\gamma}[E_i/A]$) and $\kappa_{\rm mes}[E_i]\sim \kappa_{p\gamma}[E_i/A]/A$, we approximately obtain~\cite{Murase:2010gj}
%===================================
\begin{equation}
E_\nu^2\frac{d J_\nu}{d E_\nu}\approx \frac{3}{8}\kappa_{p\gamma}\tau_{p\gamma}[E_i/A] E_i^2\frac{d J_{\rm CR}}{d E_i}
\end{equation}
%===================================
We stress that this formula is derived assuming that all UHECRs are nuclei. Then, noting that $E_i\approx A E_p$, we have  
%===================================
\begin{equation}
E_\nu^2\frac{d J_\nu}{d E_\nu}\approx 
\frac{3}{8}\kappa_{p\gamma}\tau_{p\gamma}[E_p] E_p^2\frac{d J_{\rm CR}}{d E_p}A^{2-\alpha_{\rm CR}}.
\label{eq:neutrini_intensity_from_nuclei}
\end{equation}
%===================================
Finally, the results on such a nuclear case is obtained by introducing the following ``correction'' to the proton case considered before, which is 
\begin{equation}
E_\nu^2\frac{d J_\nu}{d E_\nu}\approx E_\nu^2\frac{d J_\nu^{(p)}}{d E_\nu}A^{2-\alpha_{\rm CR}}.
\label{eq:neutrini_intensity_from_nuclei}
\end{equation}
%=========================

This is simply because a neutrino with $E_\nu$ mainly originates from nuclei with $E_A\sim20A E_\nu$.
Thus, the diffuse constraints derived because the proton composition is regarded as conservative. 

We ``require'' that the sources should be effectively transparent to the photodisintegration process, 
and the spectrum of escaping cosmic rays should be the same as that of the accelerated rays up to $E_i^{\rm max}$. 
We assume that the flux of escaping UHECRs is the same as that of the accelerated UHECRs up to $E_i^{\rm max}$
As previously discussed, the spectrum of escaping cosmic rays can be significantly different. 
This is usually expected in radiation-rich environments such as GRBs~\cite{Murase:2008mr} and blazars~\cite{Murase:2011cy}. However, 
diffuse environments such as galaxy clusters are also plausible examples~\cite{Fang:2017zjf}. 
In general, such a case requires detailed analyses but analytical formulas are
adequate for this work. 

%==========================================
\begin{figure*}
\begin{center}
\includegraphics[width=0.45\textwidth]{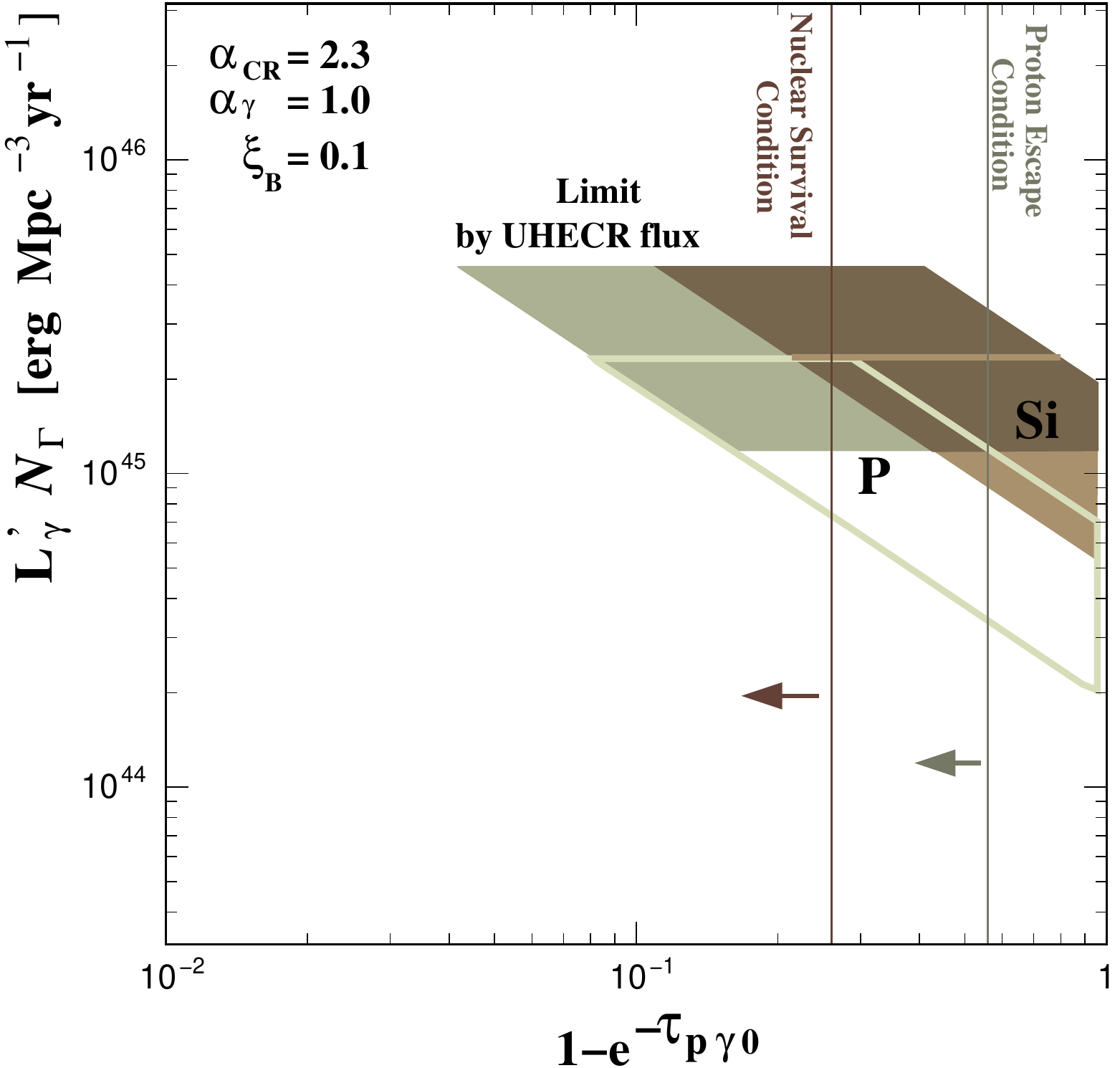}
\includegraphics[width=0.45\textwidth]{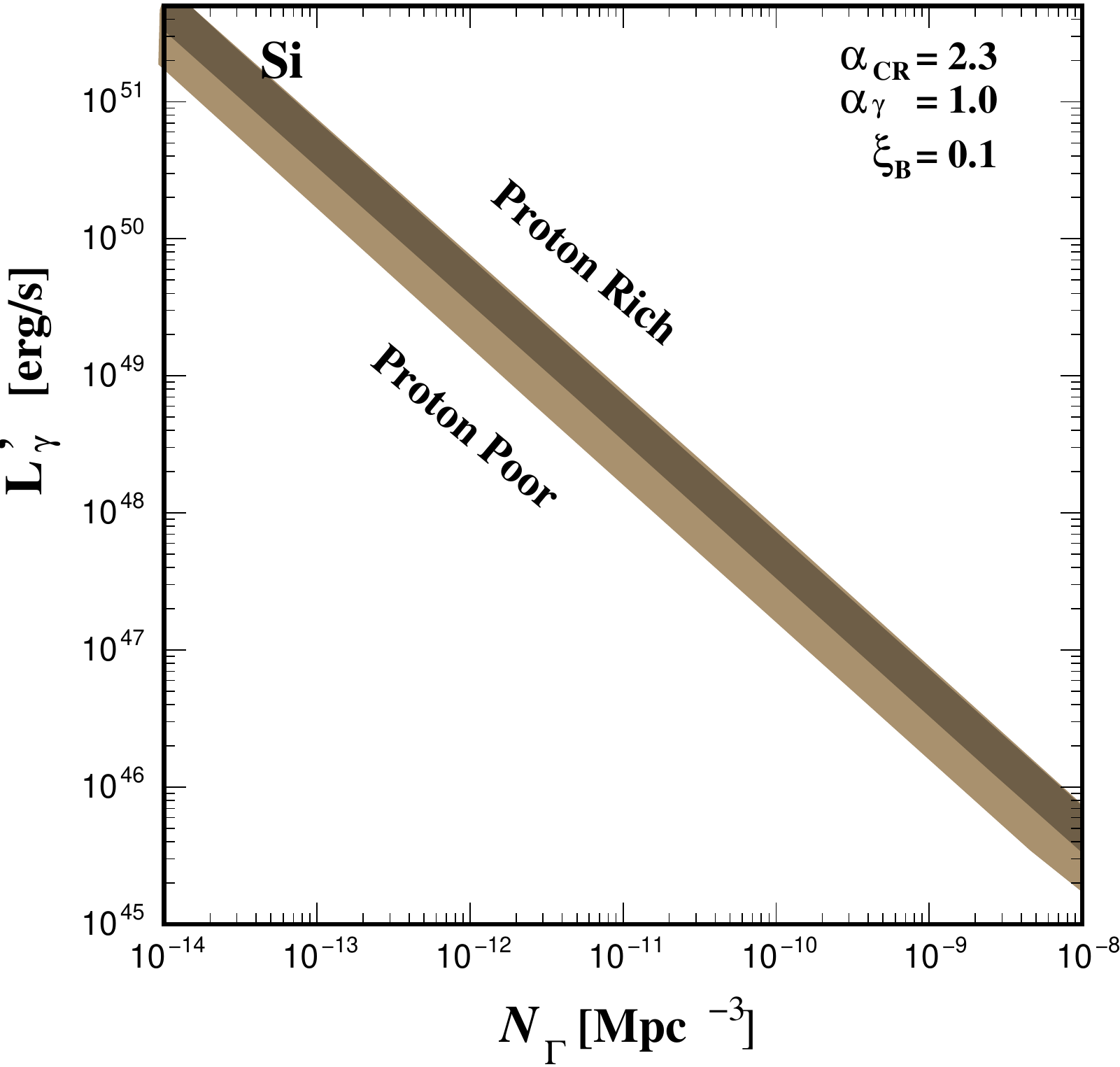}
\caption{Same as Fig.~\ref{fig:constraints_when_alpha2_2}
but show the case of  primary silicon nuclei. 
In the left plot, the constraints for the silicon case is overlaid with the proton case for comparison.
The horizontal belt represented by the darker shade shows the systematics of UHECR energetics
that originate due to the uncertainties associated with the mass composition and galactic to extra-galactic transition 
of UHECRs~\cite{Murase:2018utn}.
The darker shaded region in the right panel represents the allowed space when the nuclear-survival condition is required.}
\label{fig:constraints_nuclei}
\end{center}
\end{figure*}
%==========================================

Fig.~\ref{fig:constraints_nuclei} shows the resultant constraints (for $\alpha_\gamma=1.0$).
In this case, we consider silicon ($A=28$) UHECRs as a benchmark. 
Both the acceleration and escape conditions are considered.
The allowed region in the $L'_\gamma-N_\Gamma$ plane is similar but wider than that of the proton-dominated case. 
The allowed region for $\tau_{p\gamma0}$ is smaller in the nuclei case because of the nucleus-survival condition -- 
a photon field that facilitates the survival of nuclei is indicative of a low efficiency of photo-meson production. 
For the nucleus-survival condition, the constraints become even more stringent as indicated by the vertical line in the figure. 
This may suggest that fine-tuning is needed to build a viable model of UHECR nuclei sources.
When the target photon spectrum is softer, the resultant parameter space is even smaller
compared to the proton-dominated case.

%==================================================================================
%==================================================================================
%==================================================================================
\section{\label{sec:source}Candidate Sources}
%==================================================================================
%==================================================================================
%==================================================================================

In this section, we consider different source classes. 
The list of candidate sources for the unified photohadronic scenario is given in Table~\ref{tb1}. 

%==================================================================================
%==================================================================================
\begin{table*}[t]
\begin{center}
\caption{Characteristic parameters of the candidate sources of UHECRs and high-energy neutrinos. 
\label{tb1}
}
\scalebox{1.2}{
\begin{tabular}{|c|c|c|c|c|c|c|}
\hline         & HL GRB & LL GRB & Newborn magnetar & Jetted TDEs & Blazar Flares & Jetted AGN\\
\hline $L_\gamma$ [${\rm erg}~{\rm s}^{-1}$] & ${10}^{51-53}$ & ${10}^{46-48}$ & ${10}^{42-44}$ & ${10}^{45-48}$ & ${10}^{45-48}$ & ${10}^{43-48}$ \\
\hline $\Gamma$ & $100-1000$ & $2-30$ & $?$ & $3-100$ & $3-100$ & $3-100$ \\
\hline $\rho$ [${\rm Gpc}^{-3}~{\rm yr}^{-1}$] & $0.1-1$ & $100-1000$ & $1000-10000$ & $0.01-0.1$ & $100-1000$ & --- \\
\hline $\Delta T$ [${\rm s}$] & $10-1000$ & $100-10000$ & ${10}^{2-5}$ & ${10}^{5-7}$ & ${10}^{5-7}$ & --- \\
\hline
\end{tabular}
}
\end{center}
\end{table*}
%==================================================================================
%==================================================================================

%==================================================================================
%==================================================================================
\subsection{High-luminosity gamma-ray bursts}
%==================================================================================
%==================================================================================

%
HL GRBs are among the most powerful gamma-ray transient sources, which are classically attributed to radiation from nonthermal electrons. They are also potential candidate sources of UHECRs because of their high luminosity and large Lorentz factors~\cite{Milgrom:1995um,Waxman:1995vg,Vietri:1995hs} (see also Refs.~\cite{Murase:2008mr,Globus:2014fka,Biehl:2017zlw} for applications to nuclei). 
With $L_\gamma\sim10^{51-53}~{\rm erg}~{\rm s}^{-1}$ and $\Gamma\sim300$~\cite{Meszaros:2006rc}, we have the comoving (isotropic-equivalent) luminosity, $L_\gamma'\sim10^{46-48}~{\rm erg}~{\rm s}^{-1}$.
The magnetic energy density is assumed to be comparable to that of the radiation luminosity if the synchrotron peak is near the observed peak energy at $\varepsilon_\gamma^b\approx\Gamma\hbar{\gamma'_b}^2\frac{eB'}{m_ec}\sim300$~keV. This implies $B'\sim10^3-10^5$~G for the electron Lorentz factor $\gamma'_b\sim10^3-10^4$. This can be compatible with $\xi_B\sim0.01-100$.  
UHECR acceleration is allowed based on the luminosity argument~\cite{Murase:2008mr,Samuelsson:2018fan}. The low-energy index of the target photon spectrum (below the peak energy near $\varepsilon_\gamma^b\sim1$~MeV) is relevant for UHECRs, typically $\alpha_\gamma\sim1$, in which the photo-meson production optical depth is approximately energy independent (although multipion production enhances it by a factor of $3$~\cite{Murase:2005hy}).
%(The high-energy photon index is $\alpha_\gamma\sim2-3$, which hardens a neutrino spectrum at low energies.)
The apparent rate density of HL GRBs and the duration are $\rho\sim1~{\rm Gpc}^{-3}~{\rm yr}^{-1}$~\cite{Wanderman:2014eza} and $\Delta T\sim30$~s, respectively. This gives $n_0\sim{10}^{-15}~{\rm Mpc}^{-3}$.  
The constraint shown in Fig.~\ref{fig:constraints_when_alpha2_2} indicates that ${\mathcal N}_\Gamma\sim{10}^{-9}~{\rm Mpc}^{-3}$ (see also Figs.~\ref{fig:spectrum_example_1_0} and \ref{fig:hard_flux_scenario} for the cases with $\Gamma\sim100-1000$), which leads to $\xi_{\rm CR}\sim10{(\Gamma/300)}^{-2}$. This is consistent with the value required based on the GRB-UHECR hypothesis~\cite{Murase:2008mr}. 

One of the advantages of HL GRB models is that the steepening of neutrino spectra above a few PeV energies can readily be explained (see Fig.~2). This is because the strong cooling of pions and muons suppresses the high-energy neutrino spectrum~\cite{Waxman:1997ti}. 
However, the photo-meson production optical depth required for the unification model is $\tau_{p\gamma}\sim0.1-0.6$, which strongly constrains the HL GRB models. HL GRBs are so bright that stacking limits are very impactful, and the recent IceCube analysis gives the stringent limit, $\tau_{p\gamma}\lesssim0.05$~\cite{Abbasi:2012zw,Aartsen:2014aqy,Aartsen:2017wea} and challenges the GRB-UHECR models~\cite{Bustamante:2014oka,Bustamante:2016wpu}. 
The null detection of cosmogenic neutrinos by IceCube also substantially constrained 
the possibility that the HL GRBs are a significant population of UHECR sources~\cite{Aartsen:2016ngq}.
Thus, although the allowed parameter space may be compatible with the GRB models, we conclude that HL GRBs are unlikely to provide a unified explanation for UHECRs and PeV neutrinos.

%==================================================================================
%==================================================================================
\subsection{Low-luminosity gamma-ray bursts and transrelativistic supernovae}
%==================================================================================
%==================================================================================
%

Engine-driven supernovae with a Lorentz factor of $\Gamma\beta\gtrsim0.1-1$ have been proposed as the main sources of UHECRs~\cite{Murase:2006mm,Gupta:2006jm,Wang:2007ya,Murase:2008mr,Zhang:2018agl}. Note that this category includes LL GRBs like GRN 060218~\cite{Soderberg:2006vh}, peculiar hypernovae like SN 2009bb~\cite{Margutti:2018rri}, and fast-rising blue optical transients such as AT2018cow~\cite{Margutti:2018rri,Coppejans:2020nxp}.
If the jet scenario is assumed, with $L_\gamma\sim10^{46-48}~{\rm erg}~{\rm s}^{-1}$ and $\Gamma\sim3$, we have $L_\gamma'\sim10^{45-47}~{\rm erg}~{\rm s}^{-1}$. The luminosity requirement can be satisfied only for optimistic parameters, e.g., $L_\gamma\sim10^{48}~{\rm erg}~{\rm s}^{-1}$, but it can be more readily fulfilled if UHECRs are heavy nuclei, as opposed to considering protons ~\cite{Murase:2008mr,Samuelsson:2020upt}. 
The rate density and duration are $\rho\sim100-1000~{\rm Gpc}^{-3}~{\rm yr}^{-1}$ and $\Delta T\sim3000$~s, respectively~\cite{Campana:2006qe,Soderberg:2006vh,Liang:2006ci}, which should be compared to $N_\Gamma\sim{10}^{-9}~{\rm Mpc}^{-3}$ from Fig.~3. 
The effective number density is $n_0\sim{10}^{-11}-10^{-10}~{\rm Mpc}^{-3}$ and the condition can be satisfied if $\xi_{\rm CR}\sim(1-10){(\Gamma/3)}^{-2}$.

The peak energy of GRB 060218 and GRB 100316D is $\varepsilon_\gamma^b\sim1-10$~keV~\cite{Campana:2006qe}. 
The magnetic field strength is not well understood,
but $\xi_B\sim0.1-10$ is expected in the case of the synchrotron. For $\alpha_\gamma\sim1$, the optical depth required for photo-meson production is estimated to be $\tau_{p\gamma}\sim0.01-1$~\cite{Murase:2006mm}. It should be noted that such a hard photon spectrum is necessary to maintain consistency with optical observations~\cite{Murase:2006mm,Samuelsson:2020upt}. 
Thus, we conclude that LL GRBs could be viable sources of high-energy neutrinos and UHECRs if the luminosity is higher and/or if cosmic rays are nuclei, which is consistent with previous works~\cite{Murase:2008mr,Biehl:2017qen}. 

However, it should be considered that the mechanism of prompt emission from LL GRBs is still under debate, and another (more promising) possibility is the shock breakout scenario~\cite{Campana:2006qe}, in which gamma rays are attributed to shock breakout from a mildly relativistic outflow (that may be driven by a jet).  
In this scenario, UHECRs are unlikely to be generated during the prompt phase~\cite{Kashiyama:2013ata}. Although IceCube neutrinos are explained by choked jets or transrelativistic shocks in a dense wind~\cite{Senno:2015tsn}, UHECRs acceleration is attributed to a later transrelativistic component that is decelerated over the time scale of weeks or months ~\cite{Zhang:2017moz}.

%==================================================================================
%==================================================================================
\subsection{Newborn magnetars}
%==================================================================================
%==================================================================================

%
Some of the supernovae are more powerful than ordinary supernovae, and are referred to as hypernovae. Their ejecta are either nonrelativistic or transrelativistic (i.e., the Lorentz factor is $\Gamma\beta\gtrsim0.1-1$), which may be driven by some central engine with possible candidates that include a newborn magnetar 
(e.g.,~\cite{Thompson:2004wi}), a fallback disk around a black hole (e.g.,~\cite{Dexter:2012xk}), 
and collisions with dense circumstellar material (e.g.,~\cite{Smith:2007cb}). 

We discuss the newborn magnetar scenario that has been widely discussed in the recent literature. The spin-down luminosity is $L_{\rm sd}\sim3\times{10}^{49}~{\rm erg}~{\rm s}^{-1}$ for a millisecond rotating magnetar with a dipole magnetic field of $\sim10^{15}$~G. Efficient ion acceleration could occur inside a relativistic wind~\cite{Arons:2002yj}, in which the square of the additional factor $\theta_{\rm mag}=R_s2\pi/(cP)\sim0.2(P/1~{\rm ms})$ should be included as part of the luminosity requirement. 
Although the UHECR acceleration is possible in this magnetar scenario~\cite{Arons:2002yj}, the photons associated with the dissipation of Poynting dominated winds should be thermalized inside the supernova ejecta.  Therefore, our power-law assumption for the photon spectrum may not hold. 
Furthermore, the model typically predicts neutrino emission in the EeV range rather than in the PeV range~\cite{Murase:2009pg,Fang:2013vla}. As such, it is difficult to explain the situation of IceCube neutrinos in the PeV range using the fiducial model. Thus, this model is not discussed in further detail.
Finally, we also note that the IceCube EHE neutrino limit in the EeV range has already started to strongly constrain the magnetar scenario~\cite{Aartsen:2016ngq}.

%==================================================================================
%==================================================================================
\subsection{Tidal disruption events}
%==================================================================================
%==================================================================================

%
TDEs originate from the disruption of a main-sequence star or white dwarf by a supermassive black hole or an intermediate-mass black hole, respectively. 
Some of the TDEs have powerful jets, and the X-ray luminosity of Sw J1644+57 was $L_\gamma\sim10^{47-48}~{\rm erg}~{\rm s}^{-1}$~\cite{Burrows:2011dn}. For $\Gamma\sim10$, we have $L_\gamma'\sim10^{46-47}~{\rm erg}~{\rm s}^{-1}$. Thus the luminosity requirement can be satisfied~\cite{Farrar:2008ex}. 
The apparent rate density and duration are $\rho\sim0.01-0.1~{\rm Gpc}^{-3}~{\rm yr}^{-1}$ and $\Delta T\sim3\times{10}^6$~s, respectively. Thus the effective number density becomes $n_0\sim{10}^{-12}-10^{-11}~{\rm Mpc}^{-3}$.
In comparison to $N_\Gamma\sim{10}^{-9}~{\rm Mpc}^{-3}$ from Fig.~3, the condition for the unification of UHECRs and PeV neutrinos can be satisfied if $\xi_{\rm CR}\sim(1-10){(\Gamma/10)}^{-2}$.

The most common explanation for x rays from Sw J1644+57 is non-themnal synchrotron emission, and the peak energy is $\varepsilon_\gamma^b\approx\Gamma\hbar{\gamma'_b}^2\frac{e B'}{m_e c}\sim100$~keV~\cite{Burrows:2011dn}, which imples that $B'\sim10^2-10^4$~G for the electron Lorentz factor $\gamma'_b\sim10^4-10^5$. These can be compatible with $\xi_B\sim0.01-100$. However, provided that we consider UHECR production inside jets of TDEs such as Sw J1644+57, strong radiation fields lead to $\tau_{p\gamma0}\gg1$~\cite{Senno:2016bso}, which makes it difficult to find parameters that satisfy the constraints in Fig.~\ref{fig:constraints_when_alpha2_2}.
The problem is worse if we require the nucleus-survival condition because nuclei are disintegrated in the presence of such intense radiation fields~\cite{Zhang:2017moz,Guepin:2017abw}. It has been suggested that hypothetical low-luminosity or low-state TDEs with $L_\gamma\sim10^{45-46}~{\rm erg}~{\rm s}^{-1}$ are necessary for nuclei to survive, based on which the UHECR flux could be explained~\cite{Zhang:2017moz,Guepin:2017abw}. Alternatively, cosmic-ray acceleration at external shocks formed by jets or winds is also possible~\cite{Farrar:2014yla,Zhang:2017moz}, although efficient PeV neutrino production is not expected in these scenarios.

Our results imply that low-luminosity TDEs that allow $\tau_{p\gamma0}\lesssim1$ can satisfy the required conditions for the unification model, but nuclei rather than protons are required to obtain the highest energies. Correspondingly, the required cosmic-ray loading factors would be larger. With $N_\Gamma\sim3\times{10}^{-8}~{\rm Mpc}^{-3}$, we obtain $\xi_{\rm CR}\sim(30-300){(\Gamma/10)}^{-2}$ (see also~\cite{Zhang:2017moz,Biehl:2017hnb,Guepin:2017abw}).
However, 
%it is challenging for TDEs to be the source of a unification scenario for several different reasons. 
it is unlikely that TDEs are the common sources of IceCube neutrinos and UHECRs for several different reasons.
It has been shown that it is difficult for TDEs to be the dominant population in the diffuse IceCube flux. TDEs are so rare that the limits due to the absence of neutrino multiple sources in the IceCube data are stringent~\cite{Senno:2016bso}. Furthermore, there is no evidence of positive neutrino signals from Sw J1644+57 and other TDEs~\cite{Stein:2019ivm}. 
Recently, it has been claimed that IceCube-191001A could coincide with TDE AT2019dsg~\cite{Stein:2020xhk,Winter:2020ptf}, but the physical association is still questionable~\cite{Murase:2020lnu} although AT2019dsg is thought to be a rare, luminous class of TDEs.

%==================================================================================
%==================================================================================
\subsection{Blazar flares and jetted active galactic nuclei}%
%==================================================================================
%==================================================================================

%
Some active galactic nuclei (AGNs) have relativistic jets, and such jetted AGNs are considered as promising candidate sources of UHECRs and high-energy neutrinos. 
Recent studies have argued that steady emission of jetted AGNs is unlikely to be the source of UHECRs, especially if the UHECR composition is dominated by protons. 
Fanaroff-Riley II (FR II) galaxies and flat-spectrum radio quasars (FSRQs) can satisfy the Hillas condition, Eq.~(\ref{eq:hillas}), but they are too rare in the local universe within 100~Mpc~\cite{Takami:2008rv,Fang:2016ewe}. This difficulty can be overcome if the UHECRs are accelerated during the active/flaring phase, for which the luminosity requirement is satisfied~\cite{Murase:2008sa,Nizamov:2018sbd}.   

A typical AGN luminosity is $L_j\sim10^{44}~{\rm erg/s}$, and the isotropic-equivalent luminosity can be enhanced by $2/\theta_j^2$. 
The importance of flaring emission has been strengthened based on the recent discovery of IceCube-170922A that coincided with the flaring blazar TXS 0506+056~\cite{Aartsen2018blazar1}, although this blazar was not favored as an UHECR accelerator~\cite{Amon2018}.  

The magnetic field strength can be estimated from the Compton dominance parameter. The leptonic modeling of FSRQs often suggests $U'_\gamma\gtrsim U'_B$, and $B'\sim0.1-10$~G is typically expected for FSRQs~\cite{Ghisellini:2009fj,Murase:2014foa}, which corresponds to $\xi_B\lesssim0.01-1$. 
However, the survival of heavy nuclei is typically difficult in FSRQs, whereas low-luminosity BL Lacs allow nuclei to survive, although the photo-meson production optical depth is expected to be low~\cite{Murase:2011cy,Murase:2014foa}.
In the leptohadronic scenario (which includes the proton synchrotron scenario), higher magnetic fields, $B'\sim 10-100$~G, may be required~\cite{Petropoulou:2016ujj,Liodakis:2020dvd} but such highly magnetized environments may be highly demanding for jet physics and may be contradictory to the nucleus-survival condition (see Eq.~\ref{eq:esc_condition}).

Furthermore, UHECR emission from Fanaroff-Riley (II) galaxies/FSRQs is not favored given that strongly evolved UHECR sources are not favored by the IceCube EHE limit~\cite{Aartsen:2016ngq} as well as constraints from the absence of small-scale anisotropies. Thus, it is unlikely that the jetted AGNs are responsible for the observed UHECRs if they are dominated by protons.

Ref.~\cite{Murase:2014foa} proposed the scenario whereby EeV neutrinos are dominated by FSRQs, whereas UHECRs are dominated by BL Lac objects (see also Ref.~\cite{Rodrigues:2017fmu}).  
However, the spectrum of neutrinos is typically expected in the EeV range, so the IceCube neutrino flux is not accounted for simultaneously. The reason is as follows. 
The photo-meson production efficiency cannot decrease with the increase of energy. 
Even for FSRQs, where external radiation fields are usually dominant as target photons, $\tau_{p\gamma}$ has an energy-independent behavior beyond the pion production threshold due to the multipion production~\cite{Murase:2014foa}. 
For BL Lacs, radiation from inner jets is typically more important, and the rectangular approximation around the $\Delta$ resonance can be justified. Then, from Eq.~(\ref{eq:resnuenergy}), 1~PeV neutrinos typically originate from photons with $\sim0.8~{(\Gamma/10)}^2~{\rm keV}$. Except for extremely high synchrotron peaked BL Lacs, the spectral index in the X-ray range is around $\alpha_\gamma\sim1.5-3$, so the number of target photons is larger at lower energies. As a result, for both BL Lacs and FSRQs, the spectrum of neutrinos is predicted to be hard in the PeV range
since $\Phi_\nu \propto E_\nu^{-(\alpha_{\rm CR}+1-\alpha_\gamma)}$ as shown in Eq.~(\ref{eq:onsource_final})
(e.g.,~\cite{Mannheim:1995mm,Atoyan:2001ey,Murase:2014foa,Tavecchio:2014xha,Petropoulou:2015upa,Padovani:2015mba}
for model-dependent numerical calculations). This contradicts the diffuse limits~\cite{Aartsen:2016ngq} if the cosmic-ray spectrum is extended to ultrahigh energies with a simple power law~\cite{Dermer:2014vaa,Amon2018}. 

For example, the conclusion determined based on the model-dependent calculations for BL Lacs (that may allow the survival of nuclei) can be interpreted using Figs.~\ref{fig:hard_flux_scenario} and \ref{fig:constraints_hard_flux_scenario} considering our generic, model-independent constraints. To compensate for a soft target spectrum with $\alpha_\gamma >1$, a softer UHECR spectrum is required to supply the substantial amount of PeV energy cosmic rays as discussed in Section~\ref{subsec:parameter_relativistic}.
%sufficiently cosmic-ray spectra are required. 
For $\alpha_\gamma=1.5$ and $\alpha_{\rm CR}=2.3$, we see that the model violates the IceCube EHE limit unless $\Gamma$ is very large. Given that $\Gamma\lesssim10-100$ is expected for blazars, the cosmic-ray spectral index $\alpha_{\rm CR}$ must be larger than $2.3$ (see Fig.~\ref{fig:constraints_on_lorentz_factor}). Such cases are not excluded but the required energetics is more demanding.      
%Models explaining PeV neutrinos introduce a cutoff in the cosmic-ray spectrum~\cite{Dermer:2014vaa}.  

%==================================================================================
%==================================================================================
%==================================================================================
\section{\label{sec:summary}Summary and Discussion}
%==================================================================================
%==================================================================================
%==================================================================================

We explored the viability of the unification model for UHECRs and IceCube neutrinos considering photohadronic scenarios,
in which neutrinos are produced by the interactions between high-energy ions and low-energy photons.
The results are summarized as follows. 

%==================================================================================
\begin{itemize}
\item By requiring necessary conditions for UHECR sources,
  including those for acceleration (i.e., the Hillas condition) and survival,
  we obtained constraints on the photo-meson production optical depth in the UHECR sources.
  We further combined these source constraints with observational constraints imposed
  by the neutrino data from IceCube as well as the UHECR data from Auger.

\item We found the viable parameter space required to explain the diffuse high-energy neutrino flux
  above 100~TeV energies and the UHECR flux above 10~EeV, simultaneously.
  The allowed regions of $\tau_{p\gamma0}$ and $Q_{\rm CR}=N_\Gamma L'_\gamma$ depend on $\alpha_{\rm CR}$, $\alpha_\gamma$, and $\Gamma$.
  For $\alpha_{\rm CR}=2.2$ and $\alpha_{\gamma}=1.0$, we found $0.1\lesssim\tau_{p\gamma}\lesssim0.6$
  regardless of $\Gamma$, which can be shifted to lower values for larger $\alpha_{\rm CR}$ and/or
  smaller $\alpha_\gamma$. We also suggested the cooling break scenario,
  wherein the observed softness of the neutrino spectrum in the multi-PeV range can be explained
  by the suppression due to the cooling of mesons and muons.
  
\item The Auger data on the UHECR composition have suggested that the UHECRs are likely to be dominated
  by intermediate to heavy nuclei above the ankle. The existence of nuclei imposes an additional condition
  on their survival due to the photodisintegration process.
  We showed that the allowed parameter space is narrower than the case of only protons.
  This is mainly because the nucleus-survival condition results in tighter upper limits on the photo-meson production
  optical depth, therefore, it is more difficult for hard CR spectra and/or soft photon spectra to match the IceCube data.
  This situation is even more prominent if the observed neutrinos originate from nuclei rather than protons
  because the neutrino intensity is suppressed by $A^{\alpha_{\rm cr}-2}$ compared to the proton case
  (see Eq.~\ref{eq:neutrini_intensity_from_nuclei}).
  For example, with $\alpha_{\rm CR}\sim2.3$ and $\alpha_\gamma\sim1.0$ in the silicon composition case,
  we obtained $\tau_{p\gamma}\sim0.1\sim0.2$, which is consistent with the nucleus-survival bound
  derived by Ref.~\cite{Murase:2010gj}. The allowed parameter space is almost unique for
  $\alpha_{\rm CR}\sim2.3$ and $\alpha_{\gamma}\sim1.0$, which can be used as one of the critical tests
  for the unification model with cosmic-ray accelerators.
  
\item In general, we derived more conservative constraints that are imposed by matching the IceCube data without overshooting the Auger data.
The allowed parameter space is extended, especially for steeper cosmic-ray spectra,because larger values of the photo-meson production optical depths are possible.
It should be noted that in this case, the proton component is subdominant so UHECRs should be dominated by nuclei for a viable unification model.
  
\item Based on the conditions derived in this work, we examined different classes of astrophysical sources that could be viable as the sources of $p\gamma$ neutrinos for the unification model.
  We found that among the known source classes, LL GRBs and jetted TDEs can be viable,
  but the results of recent studies suggest that the latter source class is likely to be subdominant
  as the origin of the diffuse neutrino flux.
  However, we stress that our constraints are generic, and we do not exclude the possibility of other unknown
  source candidates. 
\end{itemize}
%==================================================================================

The grand-unification model that accounts for the gamma-ray data has been discussed,
especially for the hadronuclear scenario~\cite{Murase:2016gly,Fang:2017zjf}.
We did not explicitly calculate the extragalactic gamma-ray background that is expected in the photohadronic scenario for the unification model, because it is highly model-dependent.
In our case, gamma rays produced inside the sources are likely to be cascaded inside the sources.
There is a correspondence between the optical depth to the $\gamma\gamma\rightarrow e^+e^-$ process
and the $p\gamma$ optical depth $\tau_{p\gamma}$. Lower limits of the $p\gamma$ optical depth~\cite{Yoshida:2014uka}
suggest that it is more natural for the sources to be optically thick to GeV-TeV gamma rays~\cite{Murase:2015xka}.
However, there is an unavoidable contribution of cosmogenic gamma rays induced by UHECRs,
which can give rise to a significant contribution to the extragalactic gamma-ray background, especially
in GRB and AGN models that have strong redshift evolution.

We note that the main purpose of this work is to obtain necessary constraints for the unification model
with photohadronic neutrinos. As shown in this work, even the necessary conditions impose strict constraints,
and can allow us to determine some implications for various types of possible candidate sources.
We expect that the quantitative fitting of the data is possible but detailed analyses are left
for future work. In this case, we note that there is a large uncertainty that originates from the UHECR escape mechanism.
In the cosmic-ray accelerator models that are considered in this work,
the parameter $\alpha_{\rm CR}$ should be regarded as the spectral index of the accelerated cosmic rays,
which can be significantly different from that of the escaping UHECRs, especially
for transient sources~\cite{Zhang:2017moz,Zhang:2017hom,Zhang:2018agl}. 
As a result, the spectrum of UHECRs injected into intergalactic space
can be harder~\footnote{However, cosmic-ray reservoir models use the spectral index of cosmic rays that are injected into the environment after they escape from the sources~\cite{Fang:2017zjf}.}

%====================================================================================
%====================================================================================
\acknowledgements
%====================================================================================
%====================================================================================
The authors are grateful to Markus Ahlers and Francis Halzen
for their valuable comments on the manuscript.
This work by S.Y. is supported by JSPS KAKENHI Grant No.~18H05206 and Institute for Global Prominent
Research (IGPR) of Chiba University;
The work of K.M. is supported by the Alfred P. Sloan Foundation, NSF Grant No.~AST-1908689, and JSPS KAKENHI No.~20H01901.

%====================================================================================
%====================================================================================
%====================================================================================
\appendix

\onecolumngrid

%==================================================================================
%==================================================================================
\section{Analytical Formulas for calculating neutrino flux}
%==================================================================================
%==================================================================================
The energy flux integral, Eq.~(\ref{eq:general_t}), is transformed to
the neutrino intensity bases as
%===================================
\begin{equation}
\Phi_\nu(E_{\nu}) = \frac{c n_0}{4 \pi} 
\int_0^{\rm z_{\rm max}} 
dz \psi(z) (1 + z) \left| \frac{dt}{dz} \right| 
\frac{d\dot{N}_{\nu}}{d\varepsilon_{\nu}}(\varepsilon_{\nu}, z), 
\label{eq:general_t_flux}
\end{equation}
%===================================

The energy distribution of neutrinos generated from an interaction
that appeared in Eq.~(\ref{eq:general_yield_approx}) is given by
%===================================
\begin{equation}
Y(\varepsilon_{\nu}; \varepsilon_i, s) = \frac{1}{\sigma_{p\gamma}}\int d\varepsilon_{\pi} \frac{d \sigma_{p\gamma\rightarrow\pi}}{d\varepsilon_{\pi}}(\varepsilon_{\pi}; \varepsilon_i, s) \frac{d n_{\pi \rightarrow \nu}}{d\varepsilon_{\nu}}(\varepsilon_{\nu}; \varepsilon_{\pi}), 
\label{eq:nudist}
\end{equation}
%===================================
where $\sigma_{p\gamma\rightarrow\pi}$ is the inclusive cross-section of $p\gamma$ collisions with pion multiplicity taken into account and the last term is the neutrino spectrum from pion decay. 

Following the analytical formulation in Ref.~\cite{Yoshida:2014uka}, we finally obtain
%==================================================================================
%\begin{widetext}
\begin{equation}
\frac{d J_{\nu}}{dE_{\nu}}(E_{\nu}) \simeq 
\frac{n_0 \tau_{p\gamma 0}}{(\alpha_{\rm CR} + 1 - \alpha_\gamma)^2}\frac{K_{\rm CR}}{\varepsilon_{i0}} \frac{c}{H_0} 
\frac{s_\Delta}{\sqrt{(s_\Delta + m_{\pi}^2 - m_p^2)^2 - 4 s_\Delta m_{\pi}^2}} 
\frac{3}{1 - r_{\pi}} \left( \frac{E_{\nu}}{\varepsilon_{i0}(x_{\rm R}^+ (1 - r_{\pi}))} \right)^{-(\alpha_{\rm CR} + 1 - \alpha_\gamma)} \zeta. 
\label{eq:onsource_final}
\end{equation}
%\end{widetext}
%==================================================================================
The factor of three corresponds to the number of neutrinos produced
from the $\pi$ meson and $\mu$ lepton decay chain.
The factor $\zeta$ is the term that accounts for the redshift dependence and is given by, 
%==================================================================================
%\begin{widetext}
\begin{eqnarray}
\zeta &=& \left[I_1(z_{\rm down}, z_{\mu})+\frac{1}{3}I_1(z_{\mu}, z_{\pi}) + \right.\nonumber\\
&& \left.\frac{2}{3}\left( \frac{E_{\nu}}{\Gamma\varepsilon_{\mu}^{'\rm syn}} \right)^{-2}I_2(z_{\mu},z_{\rm max}) + 
\frac{1}{3}\left( \frac{E_{\nu}}{\Gamma\varepsilon_{\pi}^{'\rm syn}} \right)^{-2}I_2(z_{\pi},z_{\rm max})\right]\label{eq:zeta} \\
I_1(z_1, z_2) &=& \frac{2}{2 (m - \alpha_{\rm CR} + \alpha_\gamma) - 3} \Omega_{\rm M}^{- \frac{m - \alpha_{\rm CR} + \alpha_\gamma}{3}} 
\left[ \left\{ \Omega_{\rm M} (1 + z_2)^3 + \Omega_{\Lambda} \right\}^{\frac{m - \alpha_{\rm CR} + \alpha_\gamma}{3} - \frac{1}{2}} - \left\{ \Omega_{\rm M} (1 + z_1)^3 + \Omega_{\Lambda} \right\}^{\frac{m - \alpha_{\rm CR} + \alpha_\gamma}{3} - \frac{1}{2}} \right], \nonumber\\
I_2(z_1, z_2) &=& \frac{2}{2 (m - \alpha_{\rm CR} + \alpha_\gamma) - 7} \Omega_{\rm M}^{- \frac{m - \alpha_{\rm CR} + \alpha_\gamma-2}{3}} 
\left[ \left\{ \Omega_{\rm M} (1 + z_2)^3 + \Omega_{\Lambda} \right\}^{\frac{m - \alpha_{\rm CR} + \alpha_\gamma}{3} - \frac{7}{6}} - \left\{ \Omega_{\rm M} (1 + z_1)^3 + \Omega_{\Lambda} \right\}^{\frac{m - \alpha_{\rm CR} + \alpha_\gamma}{3} - \frac{7}{6}} \right].\nonumber
\end{eqnarray}
%\end{widetext}
%==================================================================================

The redshift bound $z_{\rm down}$ and $z_{\pi,\mu}$ are the bounds of the  redshift
on the UHECR sources that contribute to the neutrino flux, which are constrained by the $p\gamma$
interaction threshold and the synchrotron cooling of pion (muon), respectively. They are described as 
%===================================
%\begin{widetext}
\begin{equation}
1 + z_{\rm down} = \left\{
\begin{array}{lc}
1 + z_{\rm max} & \left( E_{\nu} < \frac{s_\Delta - m_p^2}{4 (1 + z_{\rm max})} \frac{\Gamma}{{\varepsilon'_{\gamma}}^{\rm max}} x_{\rm R}^+ (1 - r_{\pi}) \right) \\
\frac{s_\Delta - m_p^2}{4} \frac{\Gamma}{{\varepsilon'_{\gamma}}^{\rm max}} \frac{x_{\rm R}^+ (1 - r_{\pi})}{E_{\nu}} & \left( \frac{s_\Delta - m_p^2}{4 (1 + z_{\rm max})} \frac{\Gamma}{{\varepsilon'_{\gamma}}^{\rm max}} x_{\rm R}^+ (1 - r_{\pi}) \leq E_{\nu} \leq \frac{s_\Delta - m_p^2}{4} \frac{\Gamma}{{\varepsilon'_{\gamma}}^{\rm max}} x_{\rm R}^+ (1 - r_{\pi}) \right) \\
1 & \left( \frac{s_\Delta - m_p^2}{4} \frac{\Gamma}{{\varepsilon'_{\gamma}}^{\rm max}} x_{\rm R}^+ (1 - r_{\pi}) < E_{\nu} \right),
\end{array}
\right.
\label{eq:redshift_down}
\end{equation}
%\end{widetext}
%===================================
and
%===================================
\begin{equation}
1 + z_{\pi,\mu} = \left\{
\begin{array}{lc}
1 + z_{\rm max} & \left( E_{\nu} < \frac{\Gamma\varepsilon_{\nu, \pi,\mu}^{'\rm syn} }{1+z_{\rm max}} \right) \\
\frac{\Gamma\varepsilon_{\nu, \pi,\mu}^{'\rm syn} }{E_\nu} & \left(\frac{\Gamma\varepsilon_{\nu, \pi,\mu}^{'\rm syn} }{1+z_{\rm max}} \leq E_{\nu} < \Gamma\varepsilon_{\nu, \pi,\mu}^{'\rm syn}  \right) \\
1 & \left(\Gamma\varepsilon_{\nu, \pi,\mu}^{'\rm syn}  \leq E_{\nu} \right). \\
\end{array}
\right.
\label{eq:redshift_up_sync}
\end{equation}
%===================================

The third (fourth) term in the bracket in Eq.~\ref{eq:zeta} represents the spectrum
of neutrinos from synchrotron-cooled muons (pions). It should be noted that $z_{\rm down}\leq z_\mu < z_\pi$
as the synchrotron loss determines the maximal energy of neutrinos in the presented model
and $\varepsilon_{\nu,\mu}^{'\rm syn} < \varepsilon_{\nu,\pi}^{'\rm syn}$. 
$\varepsilon_{\nu,\pi/\mu}^{\rm syn}=\Gamma\varepsilon_{\nu,\pi/\mu}^{'\rm syn}$ is given by Eq.~(\ref{eq:critical_synchrotron_energy}).

$x_{\rm R}^+$ in Eqs.~\ref{eq:onsource_final} and \ref{eq:redshift_down} is the maximal bound of the relative energy of emitted pion
normalized by the parent cosmic-ray energy. They are represented by a kinematic relation
(see Eq.(6) of Ref.~\cite{Yoshida:2012gf}), 
%===================================
\begin{equation}
x_{\rm R}^{+} = \frac{(s_\Delta + m_{\pi}^2 - m_p^2) + \sqrt{(s_\Delta + m_{\pi}^2 - m_p^2)^2 - 4 s_\Delta m_{\pi}^2}}{2 s_\Delta}. 
\end{equation}
%===================================

\onecolumngrid

%==================================================================================
%==================================================================================
\section{\label{sec:UHECRspectr} Analytical formulas for estimating extragalactic UHECR intensity}
%==================================================================================
%==================================================================================

The spectrum of UHECRs injected from sources is assumed to follow a power-law form, that is
%==================================================================================
\begin{equation}
  \frac{d\dot{N}_{\rm CR}}{d\varepsilon_i}=\frac{K_{\rm CR}}{\varepsilon_{i0}}\left(\frac{\varepsilon_i}{\varepsilon_{i0}}\right)^{-\alpha_{\rm CR}}e^{-\varepsilon_i/\varepsilon_i^{\rm max}}.
  \label{eq:UHECRspec_appendix}
\end{equation}
%==================================================================================
%The superscript s represents physical quantities measured at
%a source of redshift $z_s$.

UHECRs propagate in extragalactic space and interact with CMBs via the Bethe-Heitler (BH) process
$\gamma_{\rm CMB}p\to p e^+ e^-$ and the photopion production. In the present study, we approximate that the energy attenuation length
of UHECR is constant with energies between $E_{\rm BH}$ and $E_{\rm GZK}$ governed by the BH process,
written as $\lambda_{\rm BH}$, and becomes another constant value $\lambda_{\rm GZK}$ at energies
above $E_{\rm GZK}$ where the photopion production dominates the UHECR energy loss processes.
This approximation reasonably describes the UHECR energy loss profile for the calculation of the resultant UHECR intensity
on the Earth~\cite{Takami:2007pp}, although a more accurate estimation with a precision better
than a factor of two requires dedicated numerical simulations.
We set $E_{\rm GZK}=6\times 10^{10}$~GeV and $E_{\rm BH}=2\times 10^9$~GeV, respectively.

The behaviors of UHECR propagation can then be described by classifying their energies
into five ranges for an UHECR source with a redshift of $z_s$,
(A) $\varepsilon_i<E_{\rm BH}/(1+z_s)$,
(B) $E_{\rm BH}/(1+z_s)\leq \varepsilon_i, E_i < E_{\rm BH}$,
(C) $E_{\rm BH}<E_i, \varepsilon_i \leq E_{\rm GZK}/(1+z_s)$,
(D) $E_{\rm GZK}/(1+z_s)\leq \varepsilon_i, E_i < E_{\rm GZK}$, and (E) $E_{\rm GZK}\leq E_i$.

%====================================================
%====================================================
\subsection{\label{subsec:redshift}$\varepsilon_i<E_{\rm BH}/(1+z_s)$ - The region of redshift loss only}
%====================================================
%====================================================

When the UHECR energy for a source of redshift $z_s$ is below the BH energy threshold $\varepsilon_{\rm BH}=E_{\rm BH}/(1+z_s)$,
only redshift energy loss occurs during the propagation. Given that $E_i=\varepsilon_i/(1+z_s)$,
this condition is equal to $1+z_s< \sqrt{E_{\rm BH}/E_i}$.
The UHECR spectrum from this source is given by
%==================================================================================
\begin{equation}
  \frac{d\dot{N}_{\rm CR}}{dE_i}=\frac{K_{\rm CR}}{\varepsilon_{i0}}(1+z_s)^{-(\alpha_{\rm CR}-1)}\left(\frac{E_i}{\varepsilon_{i0}}\right)^{-\alpha_{\rm CR}}.
  \label{eq:UHECRspecRedshiftOnly}
\end{equation}
%==================================================================================
Here $E_i=\varepsilon_i/(1+z_s)$ is the observed UHECR energy.

The UHECR intensity is given by
%==================================================================================
\begin{equation}
  \frac{dJ_{\rm CR}}{dE_i}=\frac{n_0c}{H_0}\int\limits^{z_{\rm UB}}_{z_{\rm LB}} dz_s \frac{\psi(z_s)}{(1+z_s)
    \sqrt{\Omega_{\rm M} (1 + z_s)^3 + \Omega_{\Lambda}}}\frac{d\dot{N}_{\rm CR}}{dE_i},
  \label{eq:UHECRintensity}
\end{equation}
%==================================================================================

where $n_0$ is the comoving UHECR source number density in the local universe and
$\psi(z_s)$ is the cosmological evolution factor of UHECR sources
and parameterized as $(1+z_s)^m$ up to $z_s=z_{\rm max}$. $z_{\rm UB}$ and $z_{\rm LB}$
are the lower and maximal bound of the source redshift distribution $z_s$, respectively.
Since $1+z_s<\sqrt{E_{\rm BH}/E_i}$, and UHECR sources are distributed between $z_s=0$ and $z_s = z_{\rm max}$,
$z_{\rm UB}$ in the integral of Eq.~\ref{eq:UHECRintensity} that is described by $z_{\rm BH}$,
which is a function of the UHECR energy on the Earth, $E_i$,
and given by

%==================================================================================
\begin{equation}
1 + z_{\rm BH} = \left\{
\begin{array}{lc}
1 + z_{\rm max} & \left( E_i < \frac{E_{\rm BH}}{(1 + z_{\rm max})^2} \right) \\
\sqrt{\frac{E_{\rm BH}}{E_i}} & \left( \frac{E_{\rm BH}}{(1 + z_{\rm max})^2}\leq E_i < E_{\rm BH} \right) \\
1 & \left( E_{\rm BH} \leq E_i \right).
\end{array}
\right.
\label{eq:redshift_bh}
\end{equation}
%==================================================================================
Obviously $z_{\rm LB} = 0$.

Putting Eq.~\ref{eq:UHECRspecRedshiftOnly} into
Eq.~\ref{eq:UHECRintensity}, we get

%==================================================================================
%\begin{widetext}
\begin{equation}
  \frac{dJ_{\rm CR}}{dE_i}\simeq n_0 \frac{K_{\rm CR}}{\varepsilon_{i0}} \frac{c}{H_0} \left(\frac{E_i}{\varepsilon_{i0}}\right)^{-\alpha_{\rm CR}}
  \frac{1}{2(m-\alpha_{\rm CR})-1}\Omega_{\rm M}^{-\frac{m-\alpha_{\rm CR}+1}{3}}
   \left[ \left\{ \Omega_{\rm M} (1 + z_{\rm BH})^3 + \Omega_{\Lambda} \right\}^{\frac{m - \alpha_{\rm CR}}{3} - \frac{1}{6}} -1\right ]
  \label{eq:UHECRintensityRedshiftOnly}
\end{equation}
%\end{widetext}
%==================================================================================
In this case, we use the approximated expression for the integral of the source redshift
described in the appendix of Ref.~\cite{Yoshida:2014uka}.

%====================================================
%====================================================
\subsection{\label{subsec:BHandRedshift}$E_{\rm BH}/(1+z_s)\leq \varepsilon_i, E_i<E_{\rm BH}$ - The region of partial BH process and redshift loss}
%====================================================
%====================================================

In this energy range, UHECR proton is subject to the redshift loss and the BH process
during its propagation until its energy reaches the threshold energy of the BH process.
Let us denote the redshift when the proton energy is equal to the BH threshold energy as
$z=\overline{z_{\rm BH}}$. In propagating from $z=\overline{z_{\rm BH}}$ to $z=0$, only the redshift dilution
reduces its energy. In this case, the UHECR proton energy on the Earth is related to $E_{\rm BH}$ by
$E_i=E_{\rm BH}/(1+\overline{z_{\rm BH}})^2$. Hence $\overline{z_{\rm BH}}$ is effectively represented
by Eq.~\ref{eq:redshift_bh}. The UHECR proton energy at source $\varepsilon_i$ is related to $E_{\rm BH}$
as
%==================================================================================
\begin{equation}
  \varepsilon_i = E_{\rm BH}\frac{1+z_s}{(1+z_{\rm BH})^2} e^{\frac{c}{H_0\lambda_{\rm BH}}\frac{2}{3\Omega_{\rm M}}\{\sqrt{\Omega_{\rm M} (1 + z_s)^3 + \Omega_{\Lambda}} -\sqrt{\Omega_{\rm M} (1 + z_{\rm BH})^3 + \Omega_{\Lambda}}\}}
\end{equation}
%==================================================================================

The UHECR spectrum from a source is given by
%==================================================================================
%\begin{widetext}
\begin{eqnarray}
  \frac{d\dot{N}_{\rm CR}}{dE_i} & =& \int d\varepsilon_i \frac{d\dot{N}_{\rm CR}}{d\varepsilon_i}\delta(E_i-\frac{E_{\rm BH}}{(1+z_{\rm BH})^2})\nonumber \\
  & = & \frac{c}{2H_0\lambda_{\rm BH}} \frac{K_{\rm CR}}{\varepsilon_{i0}}(1+z_s)^{-(\alpha_{\rm CR}-1)}\left(\frac{E_{\rm BH}}{\varepsilon_{i0}}\right)^{-\alpha_{\rm CR}}
  \frac{(1+z_{\rm BH})^{2\alpha_{\rm CR}+3}}{\sqrt{\Omega_{\rm M} (1 + z_{\rm BH})^3 + \Omega_{\Lambda}}}\nonumber\\
  && \times e^{-(\alpha_{\rm CR}-1)\frac{c}{H_0\lambda_{\rm BH}}\frac{2}{3\Omega_{\rm M}}\{\sqrt{\Omega_{\rm M} (1 + z_s)^3 + \Omega_{\Lambda}}
    -\sqrt{\Omega_{\rm M} (1 + z_{\rm BH})^3 + \Omega_{\Lambda}}\}}. \nonumber\\
\end{eqnarray}
%\end{widetext}
%==================================================================================

Given that $z_s\geq z_{\rm BH}$ in this category of $E_i$ range,
the resultant intensity of the UHECRs emitted from all sources in space is then calculated
using Eq.~\ref{eq:UHECRintensity} with $z_{\rm LB}=z_{\rm BH}$ and $z_{\rm UB}=z_{\rm max}$. We get
%==================================================================================
%\begin{widetext}
\begin{eqnarray}
  \frac{dJ_{\rm CR}}{dE_i} &\simeq& n_0 \frac{K_{\rm CR}}{\varepsilon_{i0}} \frac{c}{H_0} \left(\frac{E_{\rm BH}}{\varepsilon_{i0}}\right)^{-\alpha_{\rm CR}}
  \frac{1}{2(\alpha_{\rm CR}-1)}  \frac{(1+z_{\rm BH})^{2\alpha_{\rm CR}+3}}{\sqrt{\Omega_{\rm M} (1 + z_{\rm BH})^3 + \Omega_{\Lambda}}}
\times\nonumber\\
  && \left[ (1+z_{\rm BH})^{m-\alpha_{\rm CR}-2}-(1+z_{\rm max})^{m-\alpha_{\rm CR}-2}  e^{-(\alpha_{\rm CR}-1)\frac{c}{H_0\lambda_{\rm BH}}\frac{2}{3\Omega_{\rm M}}\{\sqrt{\Omega_{\rm M} (1 + z_{\rm max})^3 + \Omega_{\Lambda}} -\sqrt{\Omega_{\rm M} (1 + z_{\rm BH})^3 + \Omega_{\Lambda}}\}}\right ].\nonumber\\
  \label{eq:UHECRintensityBHandRedshift}
\end{eqnarray}
%\end{widetext}
%==================================================================================
In this case, we focus on the leading terms of $O(c/H_0\lambda_{\rm BH})$ after integrating the formula over $z_s$.

The UHECR intensity at an energy below $E_{\rm BH}$ is obtained based on the sum of Eqs.~\ref{eq:UHECRintensityRedshiftOnly}
and  \ref{eq:UHECRintensityBHandRedshift}.

%====================================================
%====================================================
\subsection{\label{subsec:BH}$E_{\rm BH}<E_i, \varepsilon_i\leq E_{\rm GZK}/(1+z_s)$ - The region of BH process and redshift loss}
%====================================================
%====================================================
In the energy region above $E_{\rm BH}$ but for sources with an emitted UHECR energy that is less than
the photopion production threshold energy $\varepsilon_{\rm GZK}= E_{\rm GZK}/(1+z_s)$, the
UHECR energy loss profile during the propagation is governed
by the BH process and the redshift loss during the entire path from $z=z_s$ to $z=0$.
Given that $\varepsilon_i$ is related to the UHECR energy on the Earth by
%==================================================================================
\begin{equation}
  \varepsilon_i = E_i(1+z_s) e^{\frac{c}{H_0\lambda_{\rm BH}}\frac{2}{3\Omega_{\rm M}}\{\sqrt{\Omega_{\rm M} (1 + z_s)^3 + \Omega_{\Lambda}}-1\}},
\end{equation}
%==================================================================================
the condition of $\varepsilon_i\leq E_{\rm GZK}/(1+z_s)$ can be rewritten as the boundary condition of $z_{s}$:
%==================================================================================
\begin{equation}
  \sqrt{\Omega_{\rm M} (1 + z_{s})^3 + \Omega_{\Lambda}}\leq
  1+\frac{H_0\lambda_{\rm BH}}{c}\frac{3\Omega_{\rm M}}{2}\ln\left(\frac{E_{\rm GZK}}{E_i(1+z_s)^2}\right).
  \label{eq:redshift_condition}
\end{equation}
%==================================================================================
It sets the maximal redshift of the sources that constitute the left and right-hand-side
in the preceding equation to be equal to each other.
Denoting this solution by $\overline{z_{\rm BH}^{\rm GZK}(E_i)}$,
%(\textsf{\textcolor{red}{I cannot find an analytical solution of this}}), 
the upper bound in the redshift integral of Eq.~\ref{eq:UHECRintensity}, $z_{\rm UB}$,
is described by $z_{\rm BH}^{\rm GZK}(E_i)$, which is 
is a function of $E_i$ and classified in a similar form to Eq.~\ref{eq:redshift_bh} as
%==================================================================================
\begin{equation}
1 + z_{\rm BH}^{\rm GZK} = \left\{
\begin{array}{lc}
1 + z_{\rm max} & \left( E_i < \overline{z_{\rm BG}^{\rm GZK}}^{-1}(z_{\rm max}) \right) \\
1 + \overline{z_{\rm BH}^{\rm GZK}(E_i)} & \left( \overline{z_{\rm BH}^{\rm GZK}}^{-1}(z_{\rm max}) \leq E_i <
\overline{z_{\rm BH}^{\rm GZK}}^{-1}(0)\right) \\
1 & \left( \overline{z_{\rm BH}^{\rm GZK}}^{-1}(0) \leq E_i \right),
\end{array}
\right.
\label{eq:redshift_bh_gzk}
\end{equation}
%==================================================================================
since the source redshift $z_s$ is in the range of $z_s=0$ and $z_s=z_{\rm max}$.
In this case $ \overline{z_{\rm BH}^{\rm GZK}}^{-1}$
is an inverse function that resolves $z = \overline{z_{\rm BH}^{\rm GZK}(E_i)}$.

The UHECR spectrum from a source is given by
%==================================================================================
\begin{eqnarray}
  \frac{d\dot{N}_{\rm CR}}{dE_i}&=&\frac{K_{\rm CR}}{\varepsilon_{i0}}(1+z_s)^{-(\alpha_{\rm CR}-1)}\left(\frac{E_i}{\varepsilon_{i0}}\right)^{-\alpha_{\rm CR}} \nonumber\\
  && e^{-(\alpha_{\rm CR}-1)\frac{c}{H_0\lambda_{\rm BH}}\frac{2}{3\Omega_{\rm M}}\{\sqrt{\Omega_{\rm M} (1 + z_s)^3 + \Omega_{\Lambda}}-1\}}.\nonumber\\
\end{eqnarray}
%==================================================================================

The redshift integral represented by Eq.~\ref{eq:UHECRintensity} with $z_{\rm UB}=z_{\rm BH}^{\rm GZK}$ and $z_{\rm LB}=0$
  then gives the UHECR source intensity
from sources with $z_s\leq z_{\rm BH}^{\rm GZK}$. We obtain
%==================================================================================
%\begin{widetext}
\begin{eqnarray}
  \frac{dJ_{\rm CR}}{dE_i} &\simeq& n_0\frac{K_{\rm CR}}{\varepsilon_{i0}} \lambda_{\rm BH} \left(\frac{E_i}{\varepsilon_{i0}}\right)^{-\alpha_{\rm CR}}
  \frac{1}{\alpha_{\rm CR}-1}  
  \left[ (1+\frac{m-\alpha_{\rm CR}-2}{\alpha_{\rm CR}-1}\frac{H_0\lambda_{\rm BH}}{c})\right. \nonumber\\
    &&  \left. -\{(1+z_{\rm BH}^{\rm GZK})^{m-\alpha_{\rm CR}-2}+\frac{m-\alpha_{\rm CR}-2}{\alpha_{\rm CR}-1}\frac{H_0\lambda_{\rm BH}}{c}(1+z_{\rm BH}^{\rm GZK})^{m-\alpha_{\rm CR}-5}\}  e^{-(\alpha_{\rm CR}-1)\frac{c}{H_0\lambda_{\rm BH}}\frac{2}{3\Omega_{\rm M}}\{\sqrt{\Omega_{\rm M} (1 + z_{\rm BH}^{\rm GZK})^3 + \Omega_{\Lambda}} -1\}} \right].\nonumber\\
  \label{eq:UHECRintensityBHonly}
\end{eqnarray}
%\end{widetext}
%==================================================================================
In this case, we keep terms up to the second-order for $O(c/H_0\lambda_{\rm BH})$ when we integrate the formula over $z_s$.

%====================================================
%====================================================
\subsection{\label{subsec:GZK-BH}$E_{\rm GZK}/(1+z_s)<E_i, E_i<E_{\rm GZK}$ - The region of partial GZK and BH process}
%====================================================
%====================================================
In this energy range, the UHECR energy loss profile is similar to that described in section~\ref{subsec:BHandRedshift},
but now involves photopion production.
UHECRs emitted from a source at $z=z_s$ lose their energies via photopion production until
their energies become less than the threshold energy of the photo-hadronic interactions. The BH pair production
and the redshift energy loss determines the UHECR energy profile thereafter. This transition
occurs at a redshift of $z=z_{\rm BH}^{\rm GZK}$. The UHECR energy on the Earth $E_i$ is then written as
%==================================================================================
\begin{equation}
  E_i = \frac{E_{\rm GZK}}{(1+z_{\rm BH}^{\rm GZK})^2}
  e^{-\frac{c}{H_0\lambda_{\rm BH}}\frac{2}{3\Omega_{\rm M}}\{\sqrt{\Omega_{\rm M} (1 + z_{\rm BH}^{\rm GZK})^3 + \Omega_{\Lambda}}-1\}}.
  \label{eq:UHECRenergyAfterGZK}
\end{equation}
%==================================================================================
It should be noted that $z_{\rm BH}^{\rm GZK}$ is related to the UHECR energy at a source of $\varepsilon_i$
since $\varepsilon_i$ is associated with $E_{\rm GZK}$ via the photopion production and redshift energy loss
during the propagation from $z=z_s$ to $z=z_{\rm BH}^{\rm GZK}$. It is given by
%==================================================================================
\begin{eqnarray}
  \varepsilon_i &=& E_{\rm GZK}\frac{(1+z_s)}{(1+z_{\rm BH}^{\rm GZK})^2} \nonumber\\
  && e^{\frac{c}{H_0\lambda_{\rm GZK}}\frac{2}{3\Omega_{\rm M}}\{\sqrt{\Omega_{\rm M} (1 + z_s)^3 + \Omega_{\Lambda}}-\sqrt{\Omega_{\rm M} (1 + z_{\rm BH}^{\rm GZK})^3 + \Omega_{\Lambda}}\}}. \nonumber\\
  \label{eq:UHECRenergyBeforeGZK}
\end{eqnarray}
%==================================================================================

The UHECR spectrum is calculated as
%==================================================================================
%\begin{widetext}
\begin{eqnarray}
  \frac{d\dot{N}_{\rm CR}}{dE_i} & =& \int d\varepsilon_i \frac{d\dot{N}_{\rm CR}}{d\varepsilon_i}\delta(E_i-\frac{E_{\rm GZK}}{(1+z_{\rm BH}^{\rm GZK})^2}e^{-\frac{c}{H_0\lambda_{\rm BH}}\frac{2}{3\Omega_{\rm M}}\{\sqrt{\Omega_{\rm M} (1 + z_{\rm BH}^{\rm GZK})^3 + \Omega_{\Lambda}} -1\}}) \nonumber \\
  & = & \frac{c}{H_0\lambda_{\rm GZK}} \frac{K_{\rm CR}}{\varepsilon_{i0}}(1+z_s)^{-(\alpha_{\rm CR}-1)}\left(\frac{E_{\rm GZK}}{\varepsilon_{i0}}\right)^{-\alpha_{\rm CR}}
  (1+z_{\rm BH}^{\rm GZK})^{2\alpha_{\rm CR}-2}\frac{1}{2\frac{\sqrt{\Omega_{\rm M} (1 + z_{\rm BH}^{\rm GZK})^3 + \Omega_{\Lambda}}}{(1+z_{\rm BH}^{\rm GZK})^5}
    +\frac{c}{H_0\lambda_{\rm BH}}\frac{1}{(1+z_{\rm BH}^{\rm GZK})^2}}\times \nonumber \\
&& e^{\frac{c}{H_0\lambda_{\rm BH}}\frac{2}{3\Omega_{\rm M}}\{\sqrt{\Omega_{\rm M} (1 + z_{\rm BH}^{\rm GZK})^3 + \Omega_{\Lambda}}-1\}-(\alpha_{\rm CR}-1)\frac{c}{H_0\lambda_{\rm GZK}}\frac{2}{3\Omega_{\rm M}}\{\sqrt{\Omega_{\rm M} (1 + z_s)^3 + \Omega_{\Lambda}}
    -\sqrt{\Omega_{\rm M} (1 + z_{\rm BH}^{\rm GZK})^3 + \Omega_{\Lambda}}\}}. 
\end{eqnarray}
%\end{widetext}
%==================================================================================

Using Eq.~\ref{eq:UHECRintensity} with $z_{\rm UB}=z_{\rm max}$ and $z_{\rm LB}=z_{\rm BH}^{\rm GZK}$,
the UHECR intensity is then written as 
%==================================================================================
%\begin{widetext}
  \begin{eqnarray}
    \frac{dJ_{\rm CR}}{dE_i} & =& \frac{c}{H_0} n_0\frac{K_{\rm CR}}{\varepsilon_{i0}}\left(\frac{E_{\rm GZK}}{\varepsilon_{i0}}\right)^{-\alpha_{\rm CR}}\frac{1}{\alpha_{\rm CR}-1}
  (1+z_{\rm BH}^{\rm GZK})^{2\alpha_{\rm CR}-2}\frac{e^{\frac{c}{H_0\lambda_{\rm BH}}\frac{2}{3\Omega_{\rm M}}\{\sqrt{\Omega_{\rm M} (1 + z_{\rm BG}^{\rm GZK})^3 + \Omega_{\Lambda}}-1\}}}{2\frac{\sqrt{\Omega_{\rm M} (1 + z_{\rm BH}^{\rm GZK})^3 + \Omega_{\Lambda}}}{(1+z_{\rm BH}^{\rm GZK})^5}
      +\frac{c}{H_0\lambda_{\rm BH}}\frac{1}{(1+z_{\rm BH}^{\rm GZK})^2}}\times \nonumber \\
%    && e^{\frac{c}{H_0\lambda_{\rm BH}}\frac{2}{3\Omega_{\rm M}}\{\sqrt{\Omega_{\rm M} (1 + z_{\rm BH}^{\rm GZK})^3 + \Omega_{\Lambda}}-1\}} \times \nonumber\\
&&  \left[ (1+z_{\rm BH}^{\rm GZK})^{m-\alpha_{\rm CR}-2}-(1+z_{\rm max})^{m-\alpha_{\rm CR}-2}  e^{-(\alpha_{\rm CR}-1)\frac{c}{H_0\lambda_{\rm GZK}}\frac{2}{3\Omega_{\rm M}}\{\sqrt{\Omega_{\rm M} (1 + z_{\rm max})^3 + \Omega_{\Lambda}} -\sqrt{\Omega_{\rm M} (1 + z_{\rm BH}^{\rm GZK})^3 + \Omega_{\Lambda}}\}}\right ].\nonumber\\
  \label{eq:UHECRintensityBHandGZK}
\end{eqnarray}
%\end{widetext}
%==================================================================================
In this case, we only focus on the first-order terms of $O(c/H_0\lambda_{\rm GZK})$.

The UHECR intensity at an energy between $E_{\rm BH}$ and $E_{\rm GZK}$ is obtained by the sum of Eqs.~\ref{eq:UHECRintensityBHonly}
and \ref{eq:UHECRintensityBHandGZK}.

%====================================================
%====================================================
\subsection{\label{subsec:GZK}$E_{\rm GZK}\leq E_i$ - The region of GZK process only}
%====================================================
%====================================================
UHECRs in this energy region only originate from a source within the GZK sphere of $R\lesssim \lambda_{\rm GZK}$.
$\varepsilon_i$ is related to the UHECR energy on the Earth by
%==================================================================================
\begin{equation}
  \varepsilon_i = E_i(1+z_s) e^{\frac{c}{H_0\lambda_{\rm GZK}}\frac{2}{3\Omega_{\rm M}}\{\sqrt{\Omega_{\rm M} (1 + z_s)^3 + \Omega_{\Lambda}}-1\}}.
\end{equation}
%==================================================================================

Repeating the similar described procedures, the UHECR intensity is given by
%==================================================================================
%\begin{widetext}
\begin{eqnarray}
  \frac{dJ_{\rm CR}}{dE_i} &\simeq& n_0\frac{K_{\rm CR}}{\varepsilon_{i0}} \lambda_{\rm GZK} \left(\frac{E_i}{\varepsilon_{i0}}\right)^{-\alpha_{\rm CR}}
  \frac{1}{\alpha_{\rm CR}-1}  
  \left[ 1 -(1+z_{\rm max})^{m-\alpha_{\rm CR}-2}e^{-(\alpha_{\rm CR}-1)\frac{c}{H_0\lambda_{\rm GZK}}\frac{2}{3\Omega_{\rm M}}\{\sqrt{\Omega_{\rm M} (1 + z_{\rm max})^3 + \Omega_{\Lambda}} -1\}} \right].\nonumber\\
\end{eqnarray}
%\end{widetext}
%==================================================================================

Although we obtained the analytical formula of $dJ_{\rm CR}/dE_i$, we numerically integrated the analytically obtained $d\dot{N}_{\rm CR}/dE_i$ over a source redshift in Eq.~\ref{eq:UHECRintensity}
to calculate UHECR flux in the presented study described in this report.

In the construction of the unification model, the energy density of UHECR protons
are the most relevant, rather than the detailed spectral shape, because
cosmic-ray proton emission power from sources at cosmological distances
is directly related to the observed neutrino intensity.
The applicability of the presented analytical formulation for estimating
the energy density is demonstrated by comparisons between the
estimate obtaining using the analytical formula and that obtained using a full numerical calculation.
We refer to the setup of the ``proton dip'' model in Ref.~\cite{Decerprit:2011qe}
because it facilitates straightforward comparisons. The energy density per
unit volume obtained based on robust numerical calculations for source evolution
correspond to a star formation rate
of $7.5 \times 10^{44}~{\rm erg}~{\rm yr}^{-1}~{\rm Mpc}^{-3}$ ($\alpha_{\rm CR}=2.5$)
above $10^{18}$~eV, whereas we obtained
$9.3\times 10^{44} ~{\rm erg}~{\rm yr}^{-1}~{\rm Mpc}^{-3}$.
For a stronger evolution that represents powerful radio galaxies (FR-II),
the full simulation gives 
$4.2\times 10^{44} ~{\rm erg}~{\rm yr}^{-1}~{\rm Mpc}^{-3}$ ($\alpha_{\rm CR}=2.3$)
whereas the analytical formula yields $3.4\times 10^{44}~{\rm erg}~{\rm yr}^{-1}~{\rm Mpc}^{-3}$.
We determined that the present approximated analytical formula functions
for the required precision of the generic unification model, considering
that the other uncertainties of the source are larger
than the accuracy we obtained.

%=======================================================================
%  References
%=======================================================================
\twocolumngrid
%\bibliography{kmurase,syoshida}
%\bibliography{syoshida}

\hyphenation{Post-Script Sprin-ger}

\end{document}